\documentclass{aastex}          %% The manuscript based on AASTeX v5.x
\usepackage{spr-astr-addons}    %% mimicing ASTR journal style
\usepackage{natbib}
\usepackage{graphicx}
\usepackage{multirow}
\usepackage{txfonts}
\usepackage{amssymb,amsmath,mathrsfs}
\begin{document}

\title{{\it AstroSat} Observation of Non-Resonant Type-C QPOs in MAXI J1535-571}

\shorttitle{MAXI J1535-571}
\shortauthors{Chatterjee et al.}

   \author{D. Chatterjee\altaffilmark{1,2}} \author{D. Debnath\altaffilmark{2}} \author{A. Jana\altaffilmark{2,3}} \author{J.-R. Shang\altaffilmark{4}} 
   \author{S. K. Chakrabarti\altaffilmark{2}} \author{H.-K. Chang\altaffilmark{5}} \author{A. Banerjee\altaffilmark{6}} \author{A. Bhattacharjee\altaffilmark{6}} 
   \author{K. Chatterjee\altaffilmark{2}} \author{R. Bhowmick\altaffilmark{2}} \author{S. K. Nath\altaffilmark{2}}
   \affil{email (Dipak Debnath): dipakcsp@gmail.com}
%% For single author or all the authors from an institute, use "\altaffilmark{}" only

\altaffiltext{1}{Indian Institute of Astrophysics, Koramangala, Bangalore, 560034, India.}
\altaffiltext{2}{Indian Centre for Space Physics, 43 Chalantika, Garia St. Rd., Kolkata, 700084, India.}
\altaffiltext{3}{Physical Research Laboratory, Navrangpura, Ahmedabad 380009, India.}
\altaffiltext{4}{National Space Organization, National Applied Research Laboratories, Hsinchu 30013, Taiwan.}
\altaffiltext{5}{Institute of Astronomy, National Tsing Hua University, Hsinchu 30013, Taiwan.}
\altaffiltext{6}{S. N. Bose National Centre for Basic Sciences, Salt Lake, Kolkata, 700106, India.}
%\email{debjitchatterjee92@gmail.com; dipakcsp@gmail.com; argha0004@gmail.com; chakraba@bose.res.in}
%\vs \no
%   {\small Received 2020 January 3; accepted 2020 April 5}


\begin{abstract}
Galactic transient black hole candidate (BHC) MAXI J1535-571 was discovered on 2017 September 02 simultaneously 
by {\it MAXI}/GSC and {\it Swift}/BAT instruments. It has also been observed by India's first multi-wavelength 
astronomy-mission satellite {\it AstroSat}, during the rising phase of its 2017-18 outburst. We make both the 
spectral and the temporal analysis of the source during 2017 September 12-17 using data of {\it AstroSat}'s 
Large Area X-ray Proportional Counter (LAXPC) in the energy range of $3-40$~keV to infer the accretion flow 
properties of the source. Spectral analysis is done with the physical two-component advective flow (TCAF) 
solution-based {\it fits} file. From the nature of the variation of the TCAF model fitted physical flow parameters, 
we conclude and confirm that the source was in the intermediate spectral state during our analysis period.
We observe sharp type-C quasi-periodic oscillations (QPOs) in the frequency range of $\sim 1.75-2.81$~Hz. 
For a better understanding of the nature and 
evolution of these type-C QPOs, a dynamic study of the power density spectra is done. We also investigate
the origin of these QPOs from the shock oscillation model. We find that non-satisfaction of Rankine-Hugoniot 
conditions for non-dissipative shocks and not their resonance oscillations is the cause of the observed 
type-C QPOs.
\end{abstract}

%-- ISM: jets and outflows -- radiation:dynamics}

\section{Introduction}
\label{intro}
Black holes (BHs) in binary systems emit electromagnetic radiation when the gravitational potential energy of 
the accreted matter converts into thermal energy. There exist a few BHs, 
which are transient in nature. These black hole candidates (BHCs) are very interesting to study in X-rays as 
they undergo rapid evolution in their spectral and timing properties. Different spectral states such as hard (HS), 
hard-intermediate (HIMS), soft-intermediate (SIMS), soft (SS), are generally observed during an outburst, which is 
strongly correlated with temporal properties (see, McClintock \& Remillard, 2006; Debnath et al., 2013 and references 
therein). During the outburst of a transient BHC, observed spectral states could be related to different branches of 
the so-called `q'-diagram i.e., hardness intensity diagram (HID; Belloni et al., 2005 and Debnath et al., 2008) or, 
a more physical diagram with two-component accretion rate ratio plotted against X-ray intensity (ARRID; Jana et al. 2016). 
According to Debnath et al. (2017), type-I or classical type of outbursting sources show the evolution of the spectral 
states during an outburst in the sequence: HS $\rightarrow$ HIMS $\rightarrow$ SIMS $\rightarrow$ SS $\rightarrow$ 
SIMS $\rightarrow$ HIMS $\rightarrow$ HS, while in type-II or harder type of outbursting sources SS (sometimes even SIMS) 
are missing.

Low-frequency quasi-periodic oscillations (LFQPOs) are a very common observable feature in the power density spectrum (PDS)
of stellar-mass BHs. A few BHs exhibit high-frequency QPOs in their PDSs. The frequency of the QPOs in these transient X-ray 
sources range from mHz to a few hundred Hz (Morgan, Remillard \& Greiner 1997; Paul et al. 1998; Remillard \& McClintock 2006). 
Belloni \& Hasinger (1990) first reported aperiodic variability in Cyg X-1. Low ($\sim 0.01-30$~Hz) as well as high 
($\sim 40-450$~Hz) frequency QPOs in black hole X-ray binaries (BHXRBs) are also reported by many groups of scientists 
(for a review see, Remillard \& McClintock 2006; Nandi et al. 2012; Debnath et al. 2013 and references therein). In general 
LFQPOs are classified into three types (A, B, C) based on the centroid frequency, Q-value, noise, and rms amplitude 
(Casella et al. 2005). It is observed that these LFQPOs correlated with the spectral states. Generally, 
type-C QPOs evolve monotonically and are associated with HS and HIMS, whereas type A, B QPOs are seen sporadically on and off 
in SIMS. This implies a direct connection of QPOs with the accretion flow configuration (see, Debnath et al. 2015a, 2017 
and references therein).

There are several models in the literature to explain the origin and characteristics of QPOs. LFQPOs are observed to be 
correlated with the relative dominance of the thermal and the non-thermal fluxes and state transition (Muno et al. 1999; 
Chakrabarti \& Manickam 2000; Sobczak et al. 2000; Revnivstev et al. 2000; Vignarca et al. 2003; Chakrabarti et al. 2004). 
Characteristics of trapped oscillation in a gaseous disk around a supermassive black hole ($10^{9}-10^{10}~M_\odot$)
were examined (Kato \& Fukue 1980). The obtained oscillation period was $\sim$ 100 days, a typically observed 
time variation of QSOs and Seyfert galaxies. Nowak \& Wagoner (1991) investigated the normal modes of acoustic 
oscillations within the thin accretion disks, terminated at the innermost stable orbit. Chakrabarti (1989b, 1990) showed that 
due to the strong centrifugal barrier in the vicinity of the black hole, standing axisymmetric shock could form. 
Matter puffs up in the post-shock region and forms a hot torus shape Compton cloud called the CENtrifugal pressure 
dominated BOundary Layer or CENBOL. The post-shock region could oscillate if the resonance condition is satisfied, 
i.e. the infall time scale and the cooling time scale roughly matches (Molteni, Sponholz \& Chakrabarti 1996; 
Chakrabarti et al. 2015). A quasi-periodic variation of the post-shock region (CENBOL) results in oscillation 
of the Comptonized hard X-ray intensity, which is reflected as QPO in the power-density spectra (PDS) of low time binned 
(e.g., 0.01 sec) light curves (Chakrabarti \& Manickam 2000). Molteni, Sponholz \& Chakrabarti (1996) applied 
the resonance oscillation between the bremsstrahlung cooling and infall time of the post-shock region for 
supermassive black holes. The resonance between Compton cooling and compressional heating was used to understand 
LFQPOs in stellar mass black holes (Chakrabarti \& Manickam 2000; Chakrabarti et al. 2015).

The oscillation of the transition layer of the Compton cloud due to the viscous magneto-acoustic resonance is also mentioned 
to be a possible reason for the LFQPOs (Titarchuk, Lapidus \& Muslimov 1998). The Lense-Thirring frequency due to the 
precession of a radially extended thick torus has been modeled (Ingram et al. 2009) to explain the frequency of the detected 
type-C QPOs observed in BHXRBs. According to this model, the frequency of the type-C QPO increase as the source transits 
from hard to softer spectral state along with the decrease of the outer radius of the torus. Trudolyubov et al. (1999) 
explained the evolution of LFQPOs considering the variation of the boundary between the Comptonization region and the optically 
thick accretion disk. They proposed that the QPO frequency changes due to the variation of the boundary on the viscous time 
scale. Titarchuk \& Osherovich (2000) explained the LFQPOs in terms of global disk oscillations due to the gravitational 
interaction between the compact object and the disk. This normal mode oscillation of the disk happens when the gravitational 
restoring force tries to oppose the displacement of the disk from the equatorial plane. Warped disk models subjected to 
magnetic torques are also considered to explain the LFQPOs (Shirakawa \& Lai 2002). The accretion-ejection instability 
(Tagger \& Pellat 1999) in the magnetized accretion disks are mentioned to be the reason for LFQPOs (Tagger et al. 2004). 
The oscillation of the transition layer (Nobili et al. 2000; Stiele et al. 2013) and the corona (Cabanac et al. 2010) are 
also mentioned to be the reason behind the LFQPOs. In this paper, we will try to find the origin of the observed LFQPOs from 
{\it AstroSat} LAXPC data with the shock oscillation model of Chakrabarti and his collaborators (Molteni et al. 1996; 
Ryu et al. 1997; Chakrabarti et al. 2015).

In both the rising and declining phases of a transient BHC outburst, monotonic evolutions of type-C QPOs are 
generally observed in HS and HIMS. Whereas type-A or B QPOs are observed sporadically on and off in the SIMS
(see Nandi et al. 2012). The origin of QPOs is still under debate. According to the two-component advective 
flow (TCAF) solution, the origin of three different types of QPOs is easily explained with the resonance 
oscillation (type-C), weak oscillation (type-B) of the Compton cloud boundary (i.e. shock) or mostly incoherent 
oscillation of different regions of the shock-free centrifugal barrier (type-A). 
For type-C QPOs, the shock oscillation may occur due to fulfillment of the resonance condition between the cooling 
time and infall time of the post-shock region (Molteni, Sponholz \& Chakrabarti 1996; Chakrabarti et al. 2015) or 
due to the non-fulfillment of the Rankine-Hugoniot conditions (Ryu et al. 1997). In this shock oscillation model (SOM),
the frequency of the QPOs is inversely proportional to the infall time from the location of the shock. The same shock also 
explains the spectral properties in the TCAF solution (Chakrabarti \& Titarchuk 1995; Chakrabarti 1997), and could be obtained 
directly from spectral fits with the model (Debnath et al. 2014, 2015a). TCAF model uses the property of segregation 
of the transonic flow into two components - a high angular momentum Keplerian disk component on the equatorial plane and 
a low angular momentum sub-Keplerian component or halo. 

TCAF solution has been implemented as a local additive table model in {\tt XSPEC} (Arnaud 1996) by Debnath et al. (2014) 
to fit spectra of BH candidates. The model has successfully explained the accretion dynamics of several 
BHCs (see, Debnath et al. 2015a,b; Mondal et al. 2014; Chatterjee et al. 2016, 2019, 2020).%; Chatterjee et al. 2020). 

Galactic transient MAXI J1535-571 was discovered simultaneously by {\it MAXI}/GSC (Negoro et al. 2017) and 
{\it Swift}/BAT (Kennea et al. 2017) on 2017 September 02 at sky location of RA=$15^h35^m10^s$, Dec= $-57^{\circ}10^{'}43^{''}$, 
which is near the Galactic plane. It is one of the brightest compact X-ray binaries. The source has been extensively 
monitored in multi-wavebands, starting from the radio (Tetarenko et al. 2017; Russel et al. 2017) to hard X-rays
(Negoro et al. 2017; Kennea et al. 2017; Xu et al. 2017; Shidatsu et al. 2017a,b), via optical (Scaringi et al. 2017), 
and near-infrared (Dincer 2017). The companion of the source was reported in optical (Scaringi et al. 2017) as well as in 
near-infrared (Dincer 2017). Low-frequency QPOs (LFQPOs) were reported by Mereminskiy \& Grebenev (2017). Xu et al. (2017) 
estimated the spin of the source as $a > 0.84$ indicating the black hole to be a high spinning one. Miller et al. (2018) 
estimated the spin to be $a=0.994$. Russel et al. (2017) and Tetarenko et al. (2017) reported observation of jet activity 
during the hard state of the outburst. The spectral properties of the source were analyzed by Tao et al. (2018). Stiele and 
Kong (2018) used data from {\it Swift}/XRT, {\it XMM-Newton} and {\it NICER} observatories to conduct a comprehensive timing 
and spectral analysis of this source. Baglio et al. (2018) discussed the possibility of using rapid variation in mid-IR to 
test the relation between the disc-jets. In the average power spectrum corresponding to the {\it NICER} observations, 
Stevens et al. (2018) observed an LFQPO of $\sim5.72$ Hz when the source was in SIMS. Huang et al. (2018) reported timing 
and spectral analysis using the HXMT data and proposed that the inclination of MAXI J1535-571 could be high. 

Shang et al. (2019) (hereafter Paper-I) studied the spectral and temporal properties of the source using the TCAF model 
during its initial rising phase of the outburst. {\it Swift} (XRT and BAT) and {\it MAXI} (GSC) archival data were analyzed 
using the TCAF solution to understand the accretion dynamics of the source. They observed QPOs of frequencies ranging 
from $0.44-6.48$~Hz sporadically in $15$ observations out of their studied $27$ observations of Swift/XRT. Based on the 
properties (frequency, Q-value, rms, etc.), they marked the observed QPOs as type-A, type-C. They also observed one unknown 
type (type-X) QPO of $0.44$~Hz, having a lesser Q-value (=$2.34\pm0.81$) and higher rms (=$8.59\%\pm1.56$). The spectral 
analysis was done for the initial rising phase ($\sim$ 50 days from 2017 September 4 to October 24) of the outburst 
with the TCAF model-based {\it fits} file to understand accretion flow properties of the source in details. 
Spectral states and their transitions were understood based on the nature of the evolution of the TCAF model fitted 
flow parameters. They also explained the spectral nature of the source during the entire period of the 2017-18 outburst 
based on the variation of the hardness ratios (HRs; the ratio between fluxes observed by $15-50$ keV Swift BAT with 
$2-10$ keV MAXI GSC). The entire outburst was found to be in four spectral states - HS, HIMS, SIMS, SS. 
They estimated the mass of the black hole ($M_{BH}$) to be $8.9\pm1.0~M_\odot$ based on the TCAF model fitted 
spectra. This was possible since in TCAF, $M_{BH}$ is an important model input parameter.

Sreehari et al. (2019) studied spectro-temporal properties of MAXI J1535-571 using {\it AstroSat} SXT and LAXPC10 instrument 
data in the same period of the outburst. They reported type-C QPOs with harmonics varying from $1.85-2.88$~Hz. 
They also showed that the PDS generated from the lightcurve of 3-50~keV energy range contains the fundamental QPO feature, 
where the harmonic is absent above 35~keV. They presented the spectral results with the broadband (0.7-80~keV) data using 
both phenomenological and physical models. They proposed the mass of the BH between 5.14 to 7.83 $M_\odot$. Sridhar et al. (2019) 
reported the broadband spectroscopic results of MAXI J1535-571 using {\it AstroSat} observation of one orbit. They used both SXT 
($1.3-8.0$~keV) and LAXPC10 ($3.0-70.0$~keV) data for spectral analysis with the reflection component ({\tt RELXILL} family 
of relativistic reflection models) and a general relativistic thin disk component ({\tt Kerrbb}). They estimated the mass 
and distance of the BH to be $10.39^{+0.61}_{-0.62}~M_\odot$ and $5.4^{+1.8}_{-1.1}~kpc$ respectively. Bhargava et al. (2019) 
presented the spectral and temporal results using {\it AstroSat} data of MAXI J1535-571. They found a tight correlation 
between the QPO frequencies and the photon indices. They concluded that the observed type-C QPOs formed due to the oscillation of the Compton corona. 
The spectral analysis was done using SXT ($1-8$~keV), LAXPC10 and LAXPC20 ($3.5-30$~keV) data.

In this {\it paper}, we study the 2017-18 outburst of MAXI J1535-571 using the {\it AstroSat}/LAXPC data. 
The main goal of this {\it paper} is to find the origin of the type-C QPOs as observed by {\it AstroSat} LAXPC 
from the spectral and temporal analysis under the TCAF paradigm.
We analyze the data of 62 orbits. This corresponds to six observation days (2017 September 12 to September 17) to study 
the spectral and temporal properties of the source. In Section 2, we describe the observation, data reduction, and analysis 
procedure. In Section 3, we briefly present the timing and spectral results. In Section 4, we discuss our results. 
Finally, in Section 5, we summarize our findings.

\section{Observation and data analysis}

We studied {\it AstroSat} data of Large Area X-ray Proportional Counter Unit 1 (LAXPC10) in the period of 2017 September 12-17. 
Each LAXPC payload has an effective area of $\sim 2000$~$cm^2$ at 20 keV, which is $\sim 4-5$ times higher than that of 
RXTE/PCA instruments (Antia et al. 2017). For data extraction i.e., making light curves in $1$ sec and $0.01$ sec time bins 
and spectral files, we used publicly available code (http://www.tifr.res.in/$\sim$antia/laxpc.html). We used all anode layers 
for analysis. For further analysis of the extracted light curve and spectral (.pha) files, we used {\tt HeaSoft} software 
package {\tt HEADAS 6.21} and {\tt XSPEC version 12.9.1}. For the overall study of spectral and timing properties of the source, 
we follow the analysis as described in Debnath et al. (2014, 2015a). 

{\it AstroSat} observed MAXI~J1535-571 during the initial rising phase of the 2017-18 outburst for continuous 62 orbits. 
For the timing analysis, in order to generate power-density spectra (PDS), the command {\tt powspec} of the {\tt XRONOS} 
package is used on $0.01$~sec light curves in the energy range of $3-80$~keV. The PDSs are generated considering 8192 bins per interval in 
the frequency range between 0.01 to 50 Hz. Lorentzian profiles are used to fit PDS to find the centroid frequency of the 
QPOs and the {\tt fit err} command is used to get 90\% confidence $\pm$ error limits of the model fitted parameters. 
The continuum PDSs are fitted with the combination of four Lorentzian models.

We fitted all the spectra with the current version of the TCAF model {\it fits} file as an additive table model in {\tt XSPEC}. 
We used {\tt Tbabs} as the absorption model where {\tt vern} scattering cross-section (Verner et al. 1996) was considered 
taking {\tt wilm} abundance (Wilms et al. 2000). The TCAF model input parameters are: $i)$ Keplerian disk rate ($\dot{m_d}$ 
in $\dot{M}_{Edd}$), $ii)$ sub-Keplerian halo rate ($\dot{m_h}$ in Eddington rate $\dot{M}_{Edd}$), $iii)$ location of 
the shock ($X_s$ in Schwarzschild radius $r_s$=$2GM_{BH}/c^2$), $iv)$ compression ratio ($R=\rho_+/\rho_-$, where $\rho_+$ 
and $\rho_-$ are the densities in the post- and the pre-shock flows) of the shock. In case the mass of the BH ($M_{BH}$ 
in solar mass $M_\odot$) and normalization ($N$) are not known, they have to be treated as free parameters. We used 
systematic errors of 2$\%$ for the overall fitting (see, Pahari et al. 2018; Sreehari et al. 2019; Bhargava et al. 2019). 
While fitting spectra with the TCAF model, we kept the mass of BH in the range of $7.9-9.9$~$M_\odot$ as obtained in Paper-I.

\begin{figure*}
\vskip 0.5cm
\centering
\includegraphics[width=12cm,keepaspectratio=true]{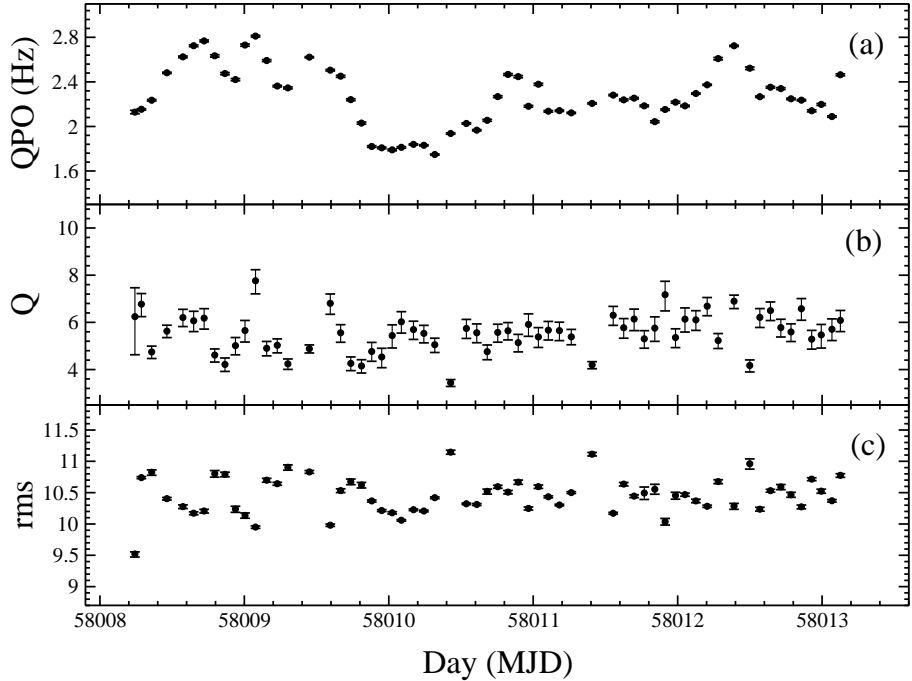}
        \caption{Variation of (a) QPO frequency (in $Hz$), (b) `Q' value, and 
	(c) percentage of $rms$ for MAXI J1535-571 using 3$-$80 $keV$ LAXPC10 data.
\label{qpo}}
\end{figure*}

\section{Results}

The spectral and temporal analysis of {\it AstroSat} LAXPC10 data during 2017 September 12 to 17 were performed. Due to the 
large effective area and data acquisition capability in low time frames (10 $\mu s$), LAXPC provides a better opportunity for 
the temporal study. Spectral data were fitted with the physical TCAF model. The analysis results are discussed in the following sub-Sections.

\subsection{Timing properties}

We searched for the QPOs in the PDSs of the fast Fourier transformed $0.01$~sec time binned light curves. The detailed nature 
of the observed LFQPOs were studied as to find their origin.

\subsubsection{Evolution of QPOs}
We found low-frequency QPOs in PDS of each of the 62 orbits data (see, Fig.~\ref{qpo}a). The QPO frequencies were observed 
to vary between $1.75-2.81$~Hz. The PDSs also showed harmonics. The centroid frequency of the QPO, `Q' value and fractional 
rms are shown in Fig.~\ref{qpo}(a-c). We checked Pearson linear correlation between the QPO frequencies and the fractional 
rms of the 62 observations. We found a very weak negative correlation between them with a correlation coefficient, r=-0.007.
Six PDSs are presented in Fig.~\ref{pds}, where the corresponding QPO frequencies are written at the upper right corner 
of each panel. In Table~\ref{table1}, we have shown the QPO properties. There was little variation in the QPO frequencies, 
but no monotonic increase or decrease in centroid frequency was observed. From the properties of the QPOs (centroid frequency, 
`Q' value, and fractional rms) observed during our studied period, we marked them as type-C QPOs. 
%However, according to Paper-I, during the LAXPC observation time, the source already entered the SIMS. The observation of type-C QPOs in SIMS is quite 
%uncommon. 
This non-evolving nature of the QPO frequency is consistent with the nature of QPOs observable in intermediate states (for more 
details see, Nandi et al. 2012; Debnath et al. 2013 and reference therein).

\subsubsection{Dynamic study of PDS}
\begin{figure*}
\vskip -0.2cm
\centering
\vbox{
\includegraphics[width=5.0truecm]{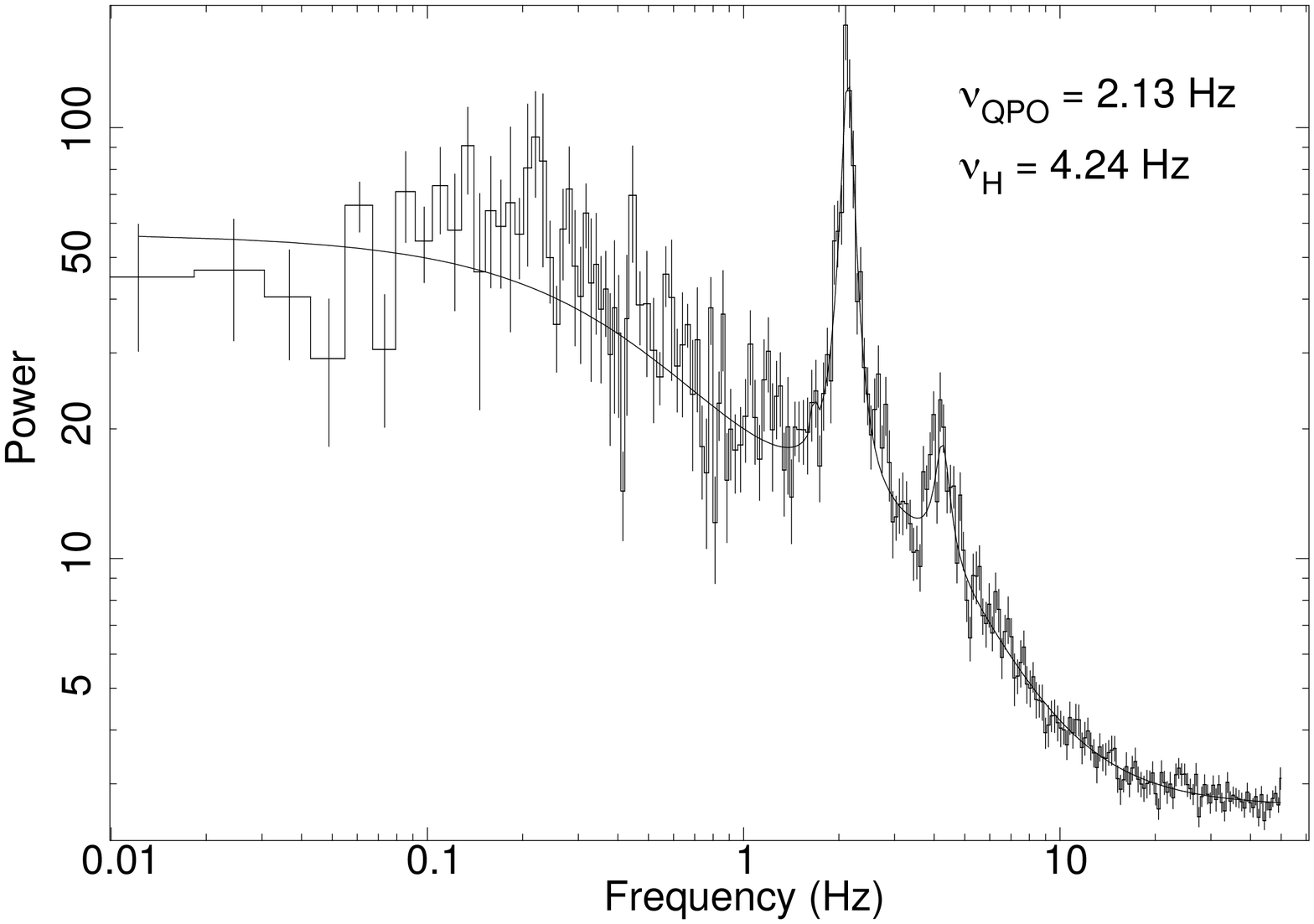}\hskip 0.5cm
\includegraphics[width=5.0truecm]{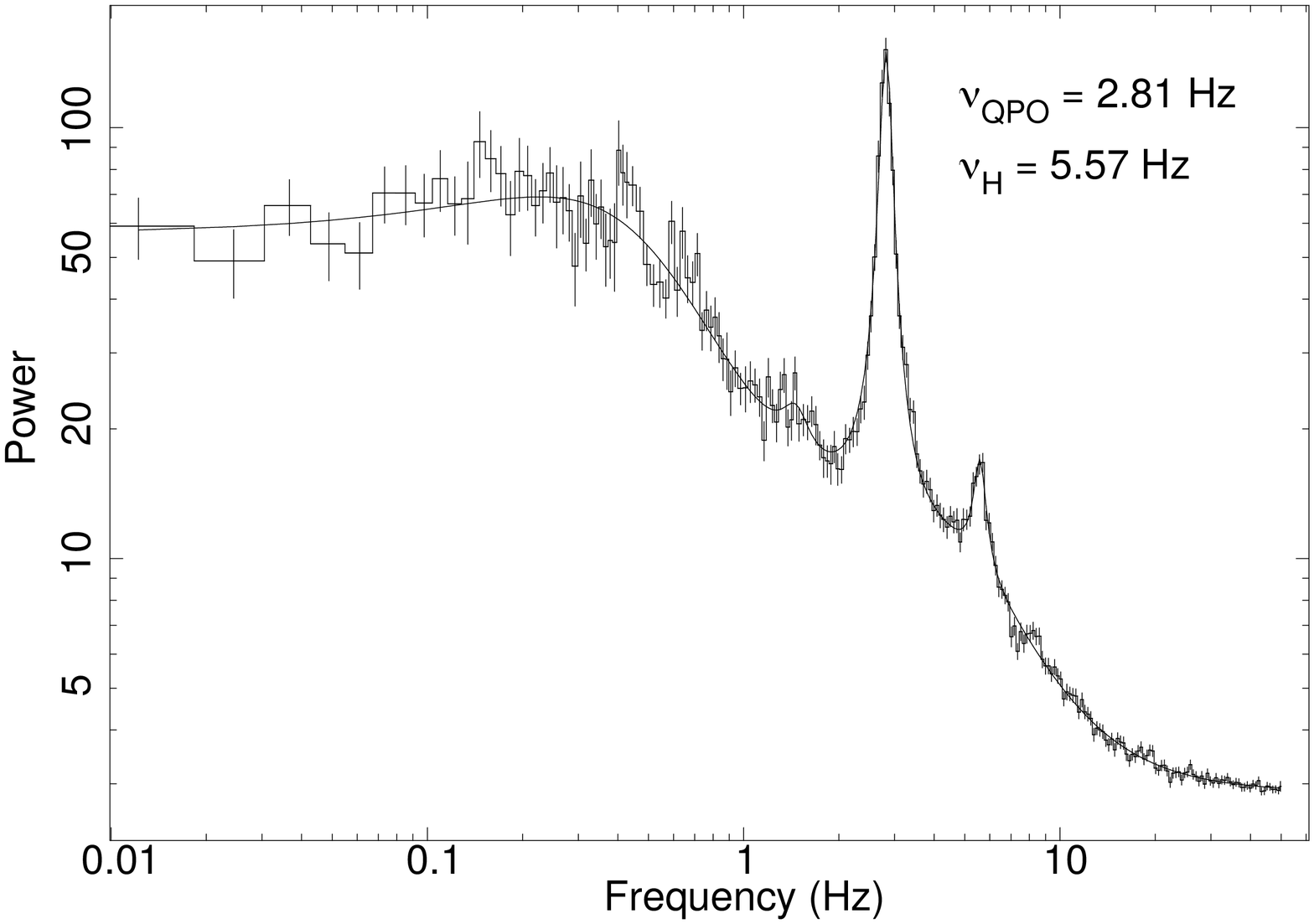}%\hskip 0.5cm
}
\vbox{
\includegraphics[width=5.0truecm]{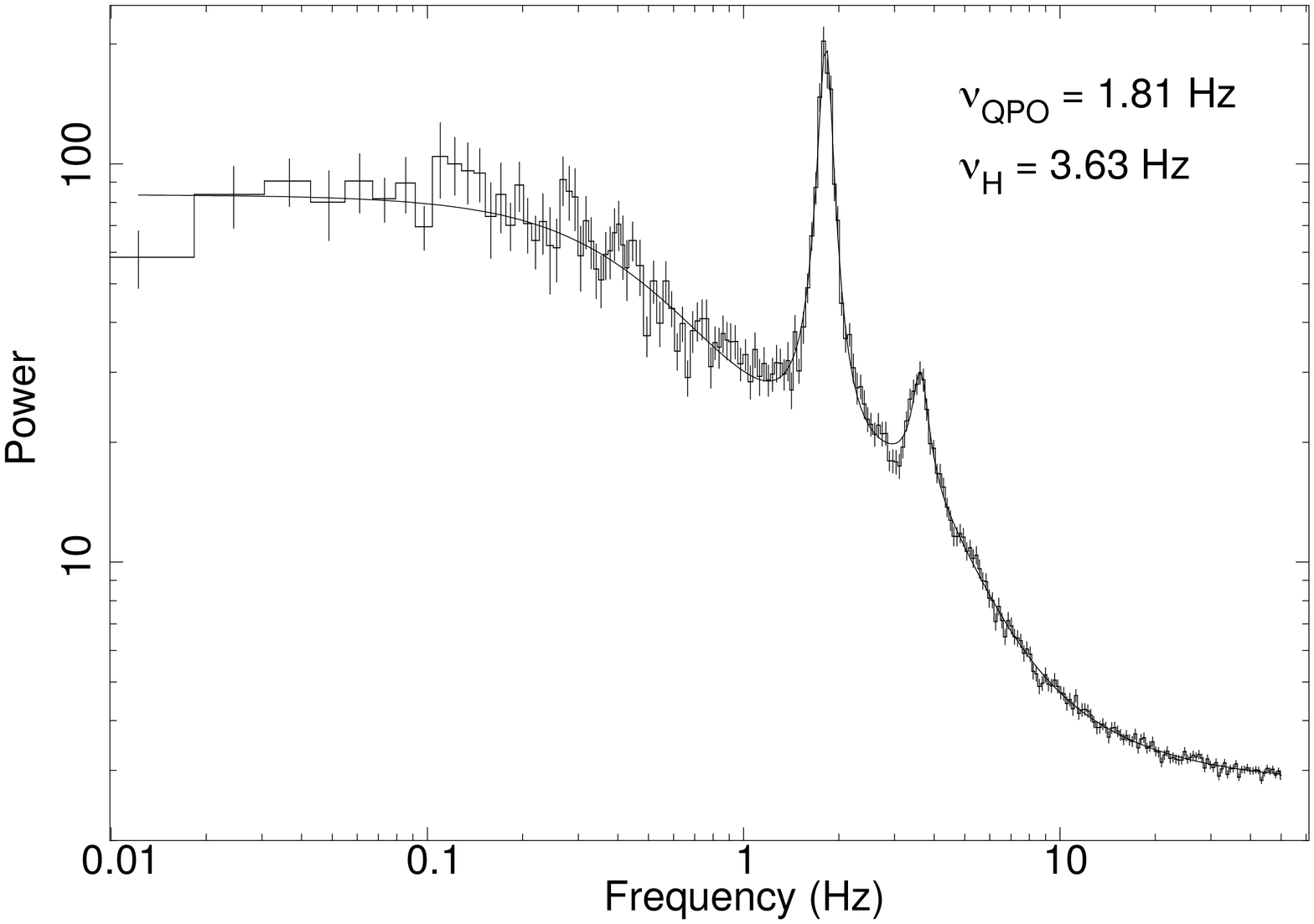}\hskip 0.5cm
\includegraphics[width=5.0truecm]{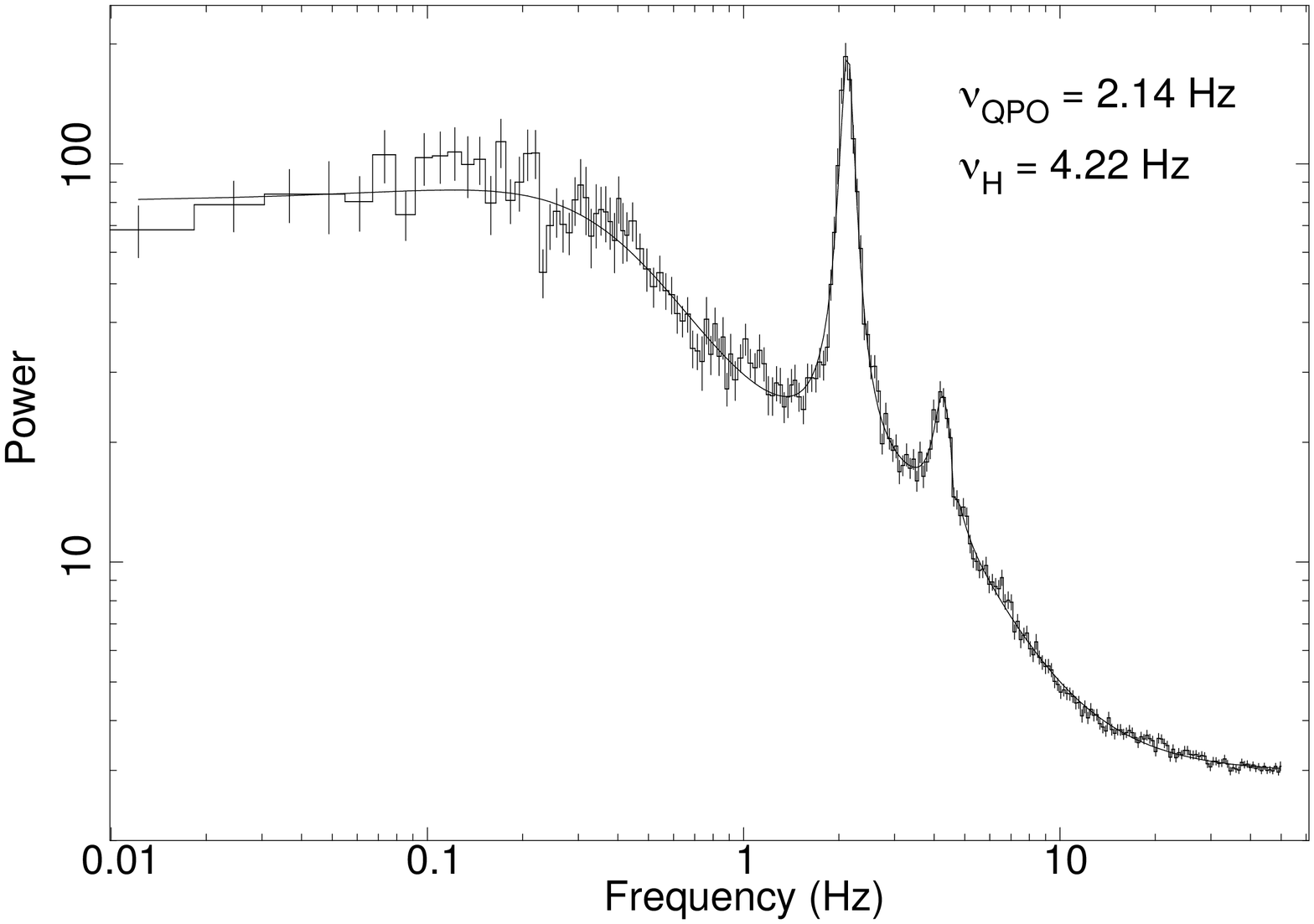}%\hskip 0.5cm
}
\vbox{
\includegraphics[width=5.0truecm]{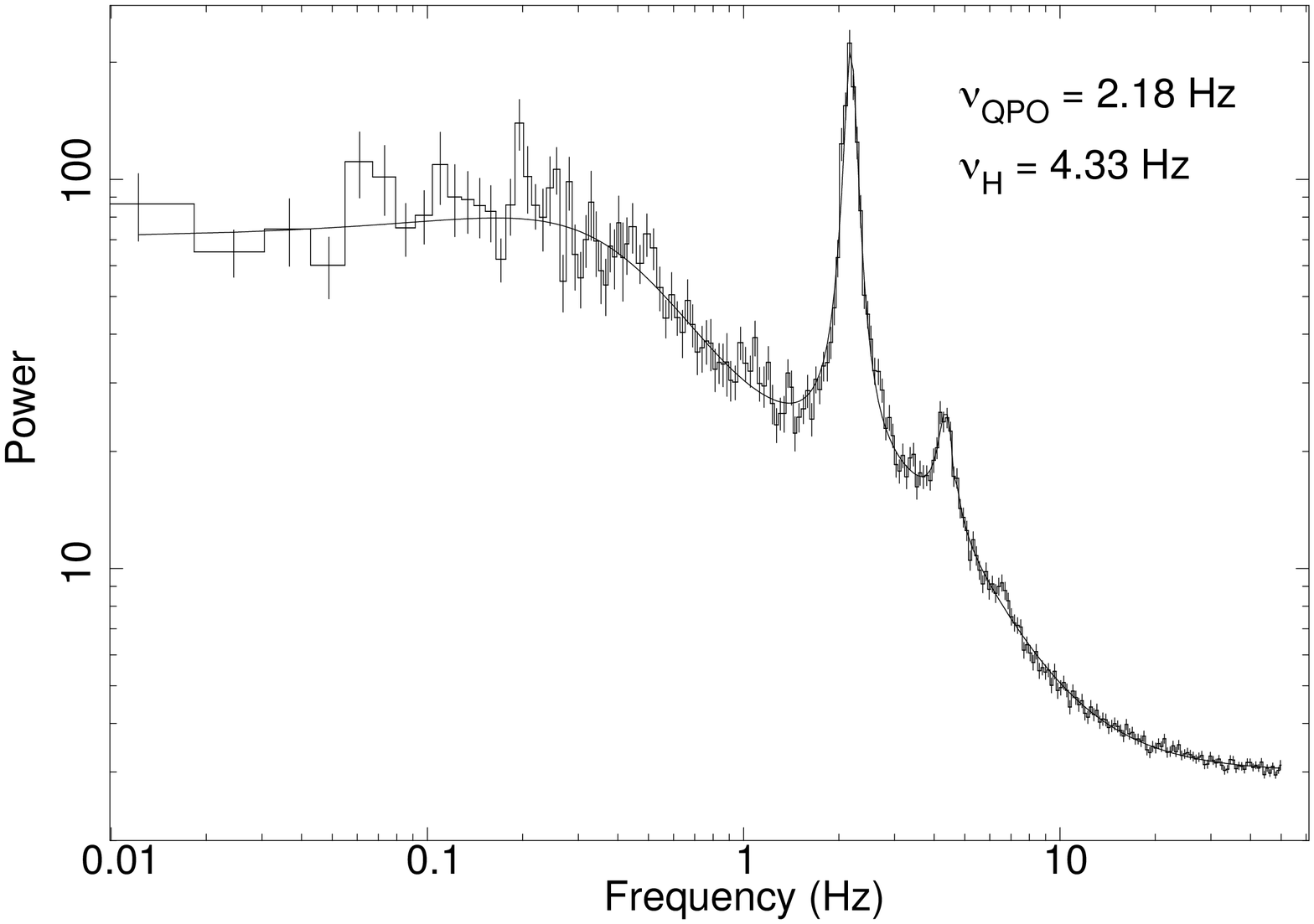}\hskip 0.5cm
\includegraphics[width=5.0truecm]{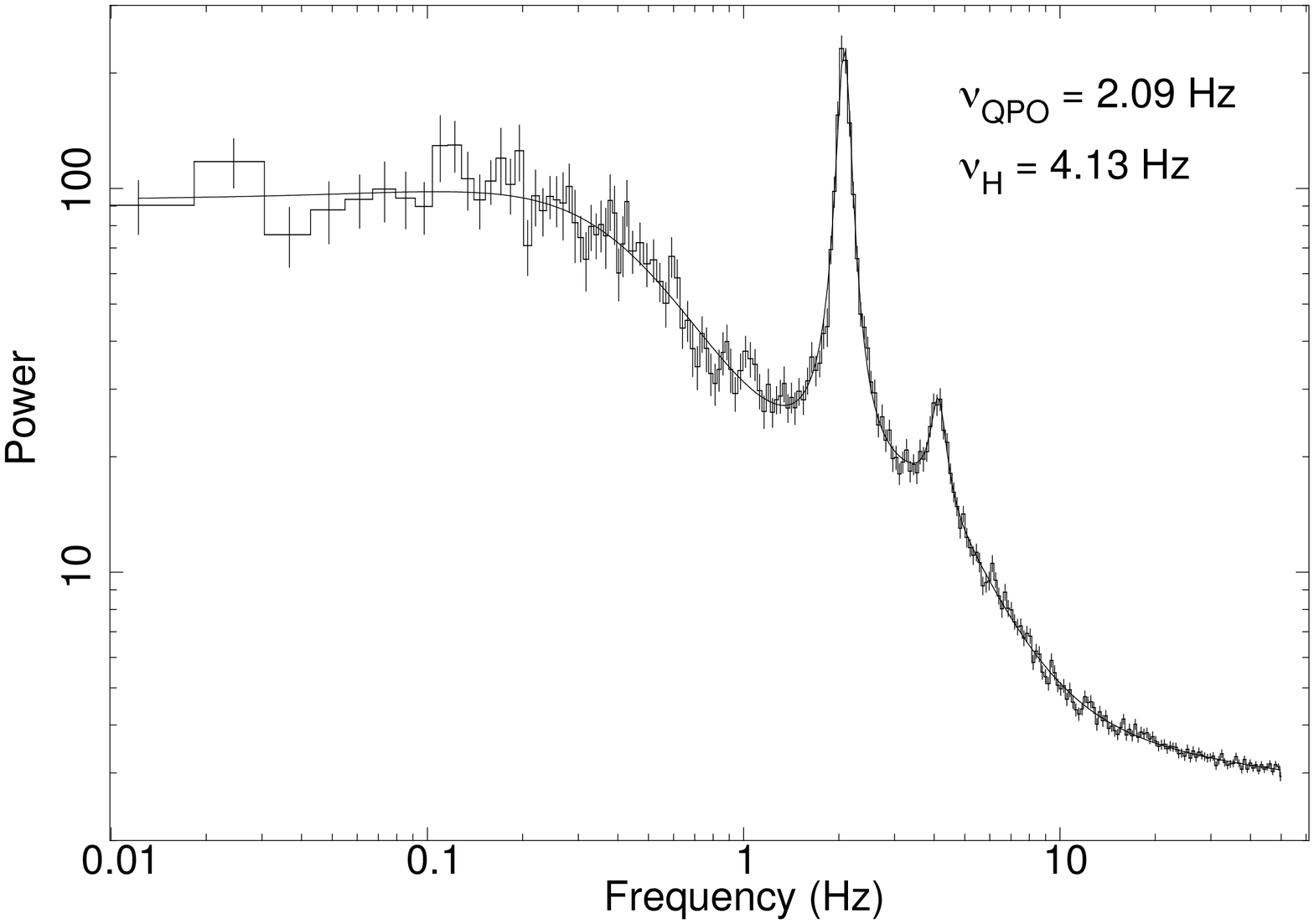} }
\vskip 0.1cm
	\caption{Leahy normalized power density spectra (PDSs) extracted from the FFT of 0.01~$sec$ time-binned light curves 
	of LAXPC10 data of MAXI J1535-571 in the frequency range between 0.01 to 50 Hz.
	The PDSs are extracted from the whole data of each orbit.
	The Lorentzian model fitted frequency of the QPOs and harmonics are marked inset in each plots.
	First panel: from left to right: orbit 10584 (12/09/2017), orbit 10597 (13/09/2017).
	Second panel: from left to right: orbit 10612 (14/09/2017), orbit 10627 (15/09/2017).
	Third panel: from left to right: orbit 10641 (16/09/2017), orbit 10656 (17/09/2017).
	\label{pds}}
\end{figure*}

\begin{figure*}
\vskip -0.5cm
\centering
\includegraphics[width=15cm,keepaspectratio=true]{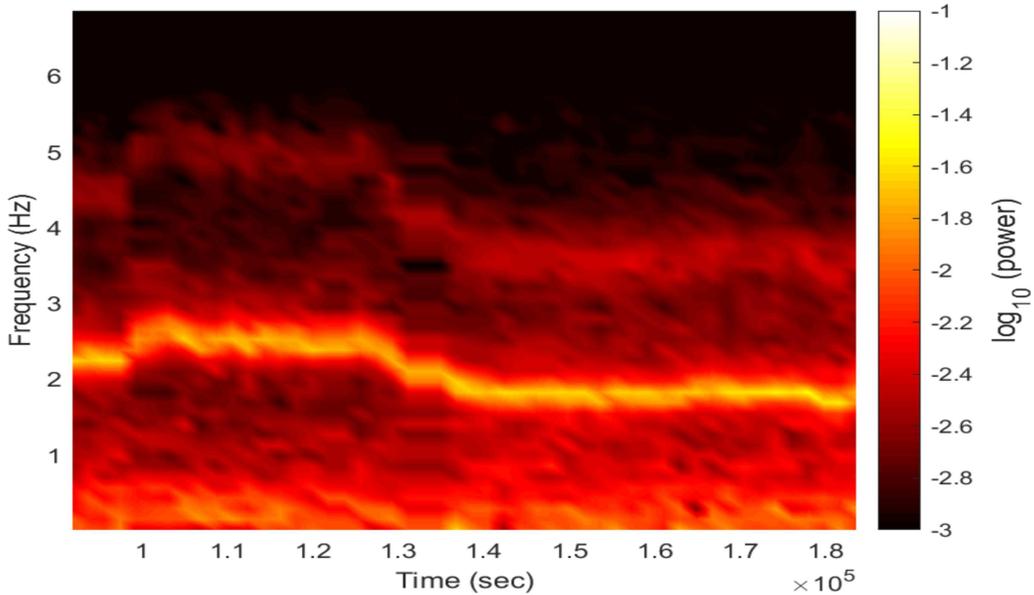}
\caption{Dynamic PDS of MAXI J1535-571 using {\it AstroSat} LAXPC10 data during the period of MJD = 58009.259 - 58010.359. 
A total of 13 orbits (10600, 106003, 10604, 10606, 10607, 10608, 10609, 10610, 10611, 10612, 10613, 10614, 10615)
data with an exposure of 91944 sec are used. The data was divided into 500 sec lightcurves with 50 sec increment.
The PDSs are extracted from those 500 sec lightcurves with 0.01 sec time bin.
\label{dynamic_pds}}
\end{figure*}

To understand the detailed nature of the variation of the observed QPOs (both primary and harmonics) during {\it AstroSat} observed 
short period of the outburst of MAXI~J1535-571, dynamic PDS is studied. Since overall fluctuation of the QPOs is not prominent 
in a small time window of each orbit data, we combined $13$ successive orbits (no. 10600$-$10615) data corresponding to 
MJD = 58009.259 to 58010.359. Total exposure time of these 13 orbits (after removing overlap time data) is 91944~sec i.e., 
$\sim$1.1~day. During this observation period, the primary dominating type-C QPO frequency was observed to vary in a very 
small range of $1.75-2.61$ Hz (see, Fig.~\ref{qpo}a and Table~\ref{table1}). A faint harmonic was also seen in the PDSs 
(see, Fig.~\ref{pds}) when we analyzed each whole orbit's data. The dynamic PDS of the combined orbits data is shown in 
Fig.~\ref{dynamic_pds}. To make this dynamic PDS, $0.01$~sec time-binned lightcurves of these orbits were divided into chunks 
of $500$~sec with an increment of $50$~sec. In Fig.~\ref{dynamic_pds}, the X-axis shows the midpoint of the time windows 
(in sec) selected for generating the dynamic PDS. The Y-axis shows the Frequency (in Hz) and the color represents 
log$_{10}$Power. The variation of centroid frequency of QPOs along with the harmonics are clearly observable in 
Fig.~\ref{dynamic_pds}.

Initial $\sim 3.7$~hrs, QPOs were roughly constant at $\sim 2.3$~Hz, then it moved to the higher frequency branch 
$\sim 2.6$~Hz. A small variation of frequency in this higher branch was observed for $\sim 7.8$~hrs. Then it moved back 
to the lower frequency branch of $\sim 2.0$~Hz, after spending small duration in $\sim 2.2$~Hz. It continued in this 
low-frequency range for the remaining time of $\sim 13.9$~hrs.

\subsection{Spectral properties}
\label{spec_prop}
\begin{figure*}
\vskip -0.1cm
\centering
\includegraphics[width=8cm,keepaspectratio=true]{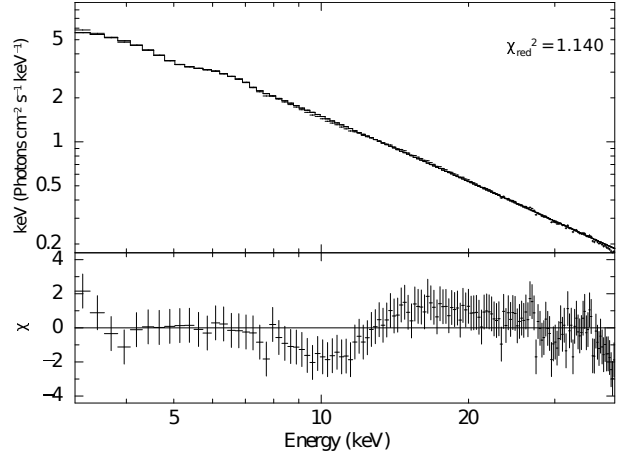}
\caption{TCAF model fitted spectrum of orbit no. 10597, observed on 2017 September 13 (MJD=58009.128). A systematic error 
of 2$\%$ was used (Sreehari et al. 2019) for LAXPC data. Note, for the same data set, Bhargava et al. (2019) used 3\% 
systematic error.
\label{tcaf_spec}}
\end{figure*}

\begin{figure*}
\vskip -0.2cm
\centering
\vbox{
\includegraphics[width=6.0truecm]{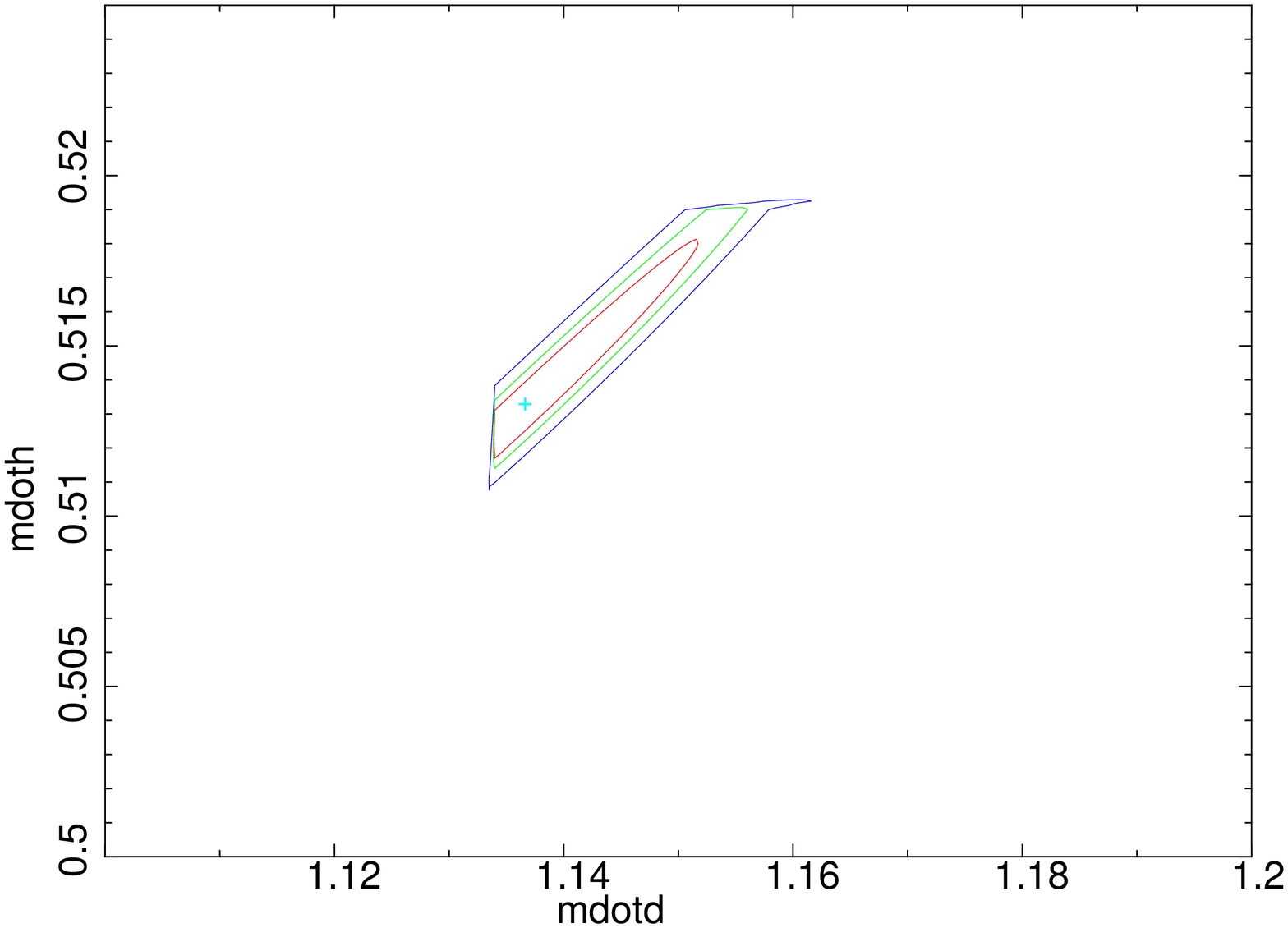}\hskip 0.5cm
\includegraphics[width=6.0truecm]{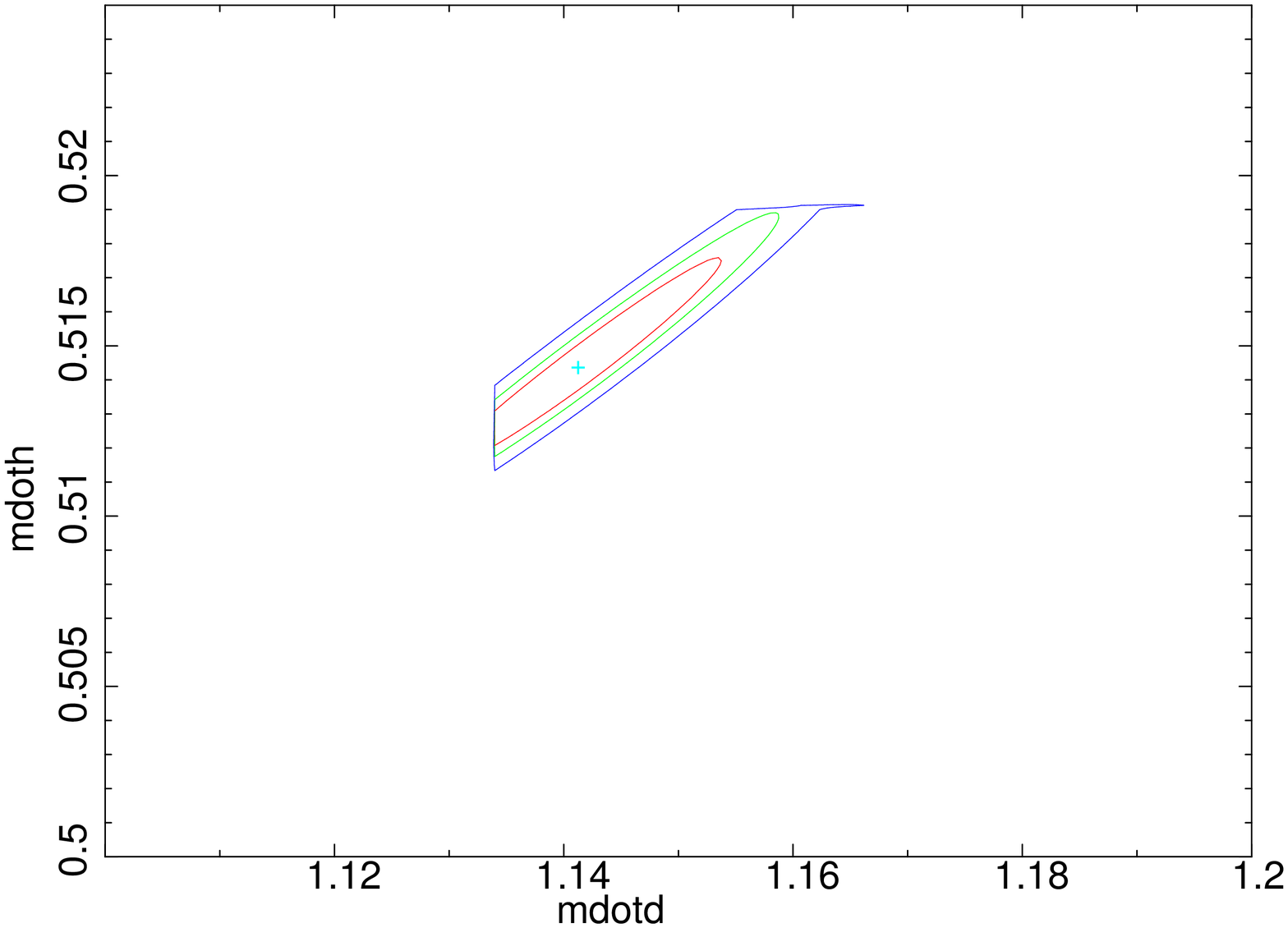}%\hskip 0.5cm
}
\vbox{
\includegraphics[width=6.0truecm]{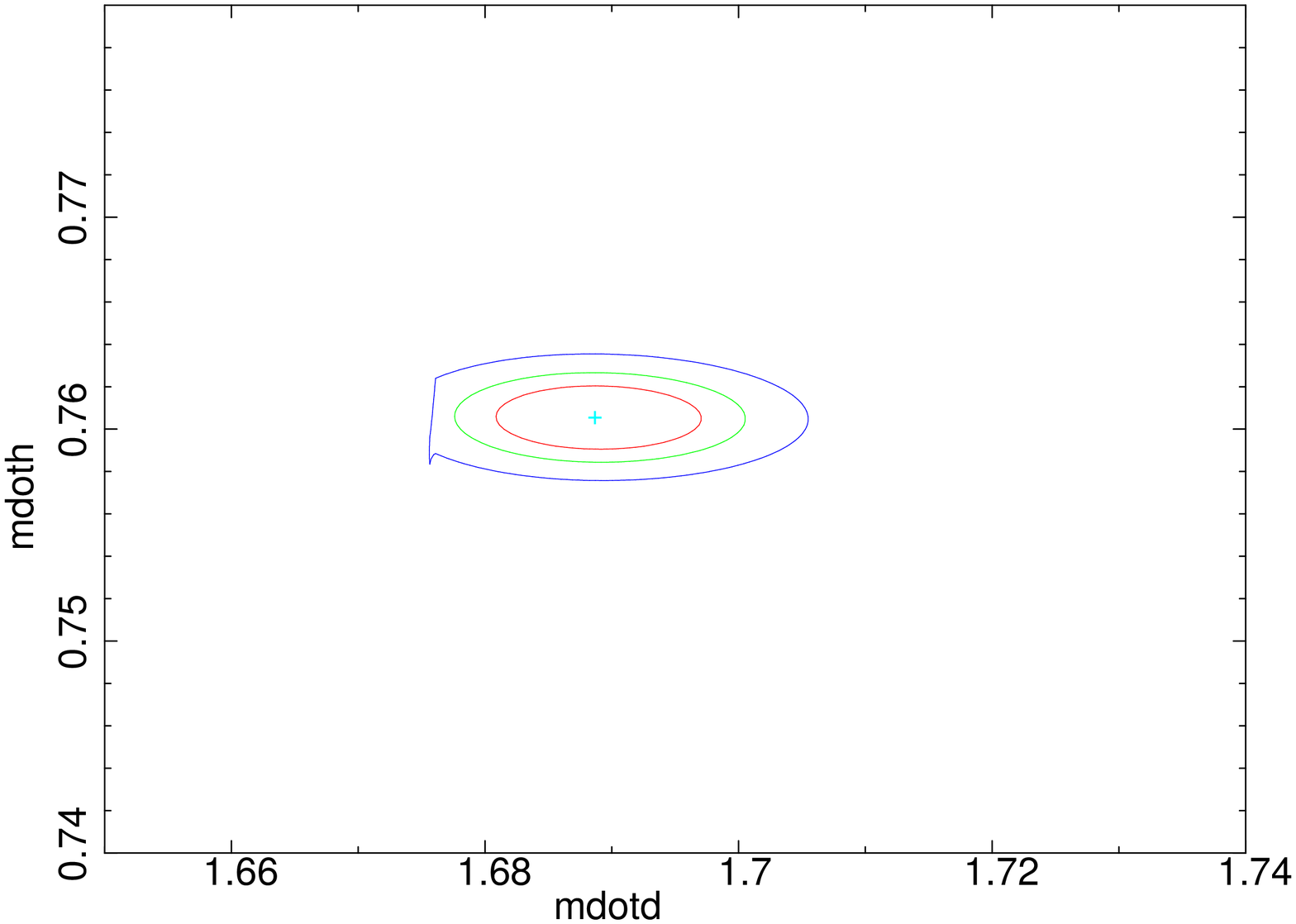}\hskip 0.5cm
\includegraphics[width=6.0truecm]{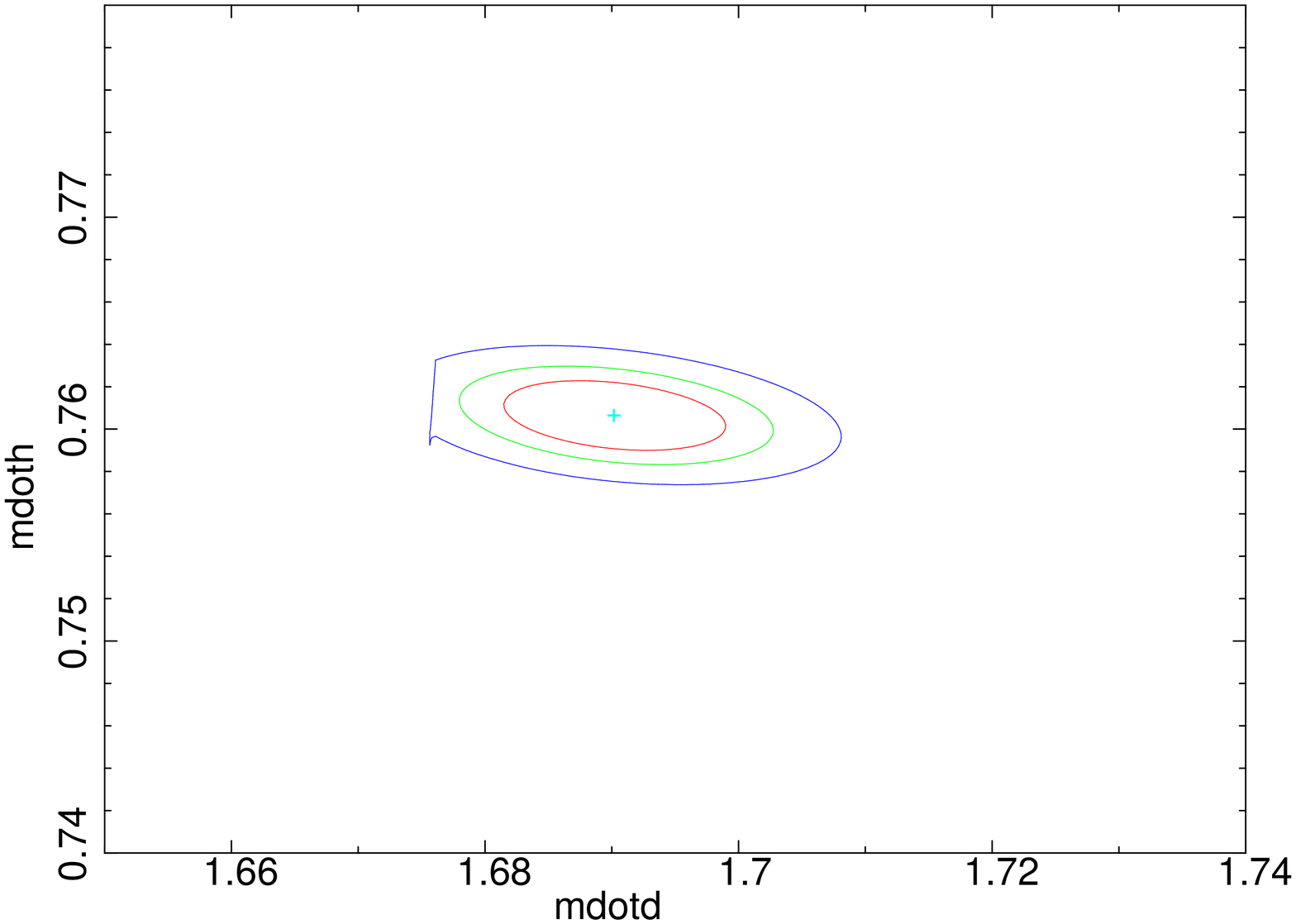}
}
\vbox{
\includegraphics[width=6.0truecm]{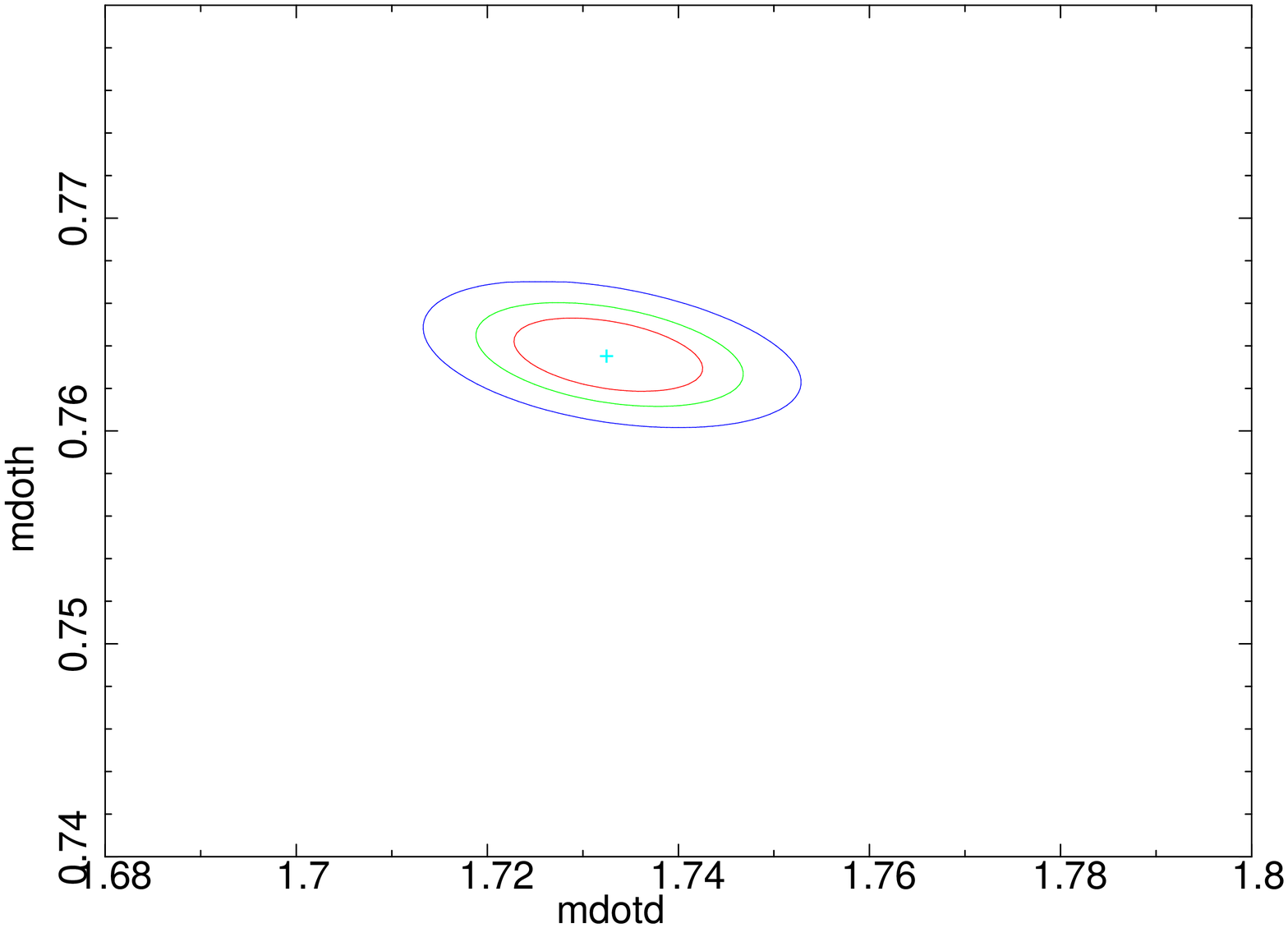}\hskip 0.5cm
\includegraphics[width=6.0truecm]{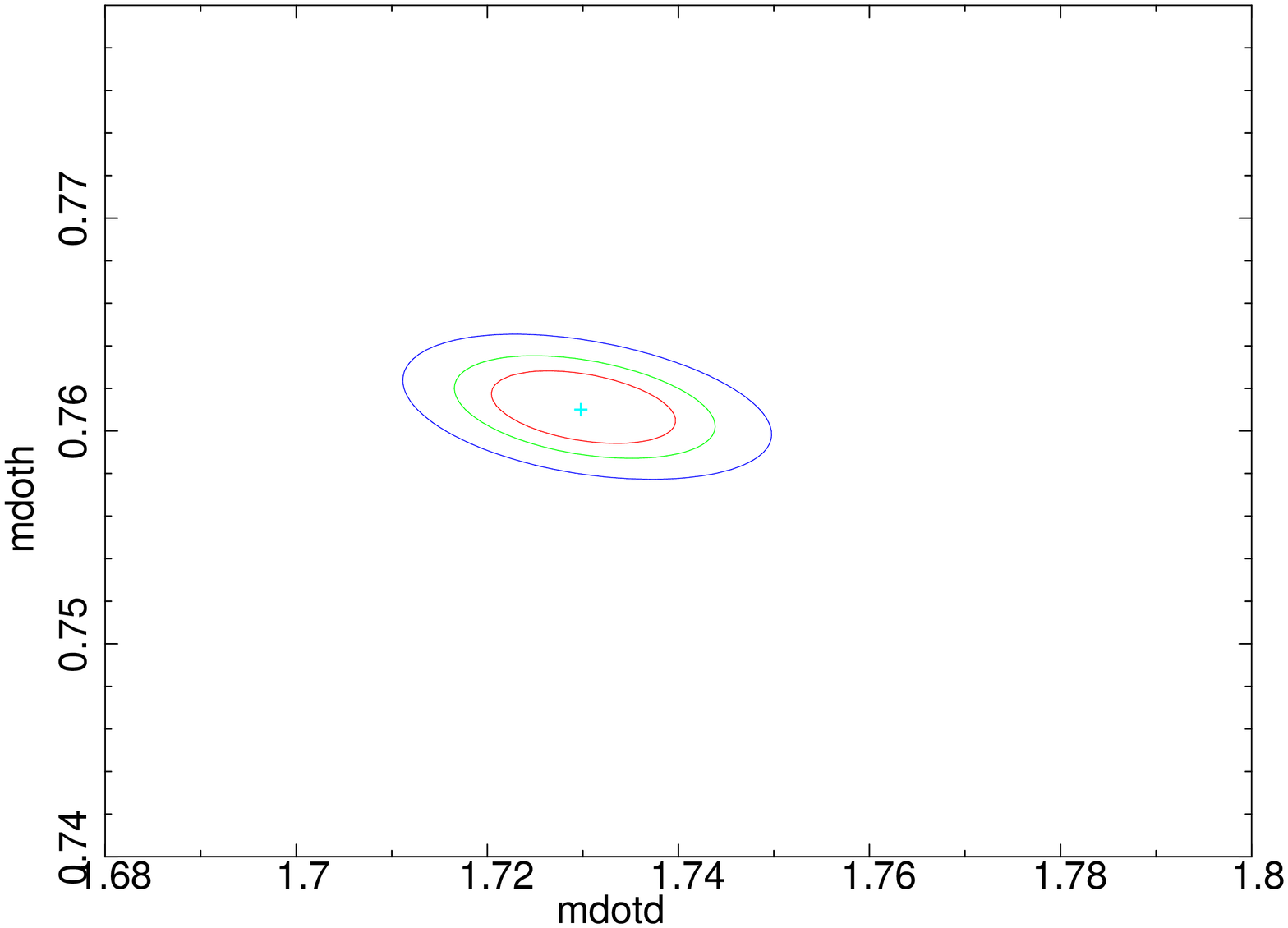} }
\vskip 0.1cm
	\caption{Confidence contours of $\dot m_d$-$\dot m_h$ of six orbits are shown.
	First panel: from left to right: orbit 10584 (12/09/2017), orbit 10597 (13/09/2017).
	Second panel: from left to right: orbit 10612 (14/09/2017), orbit 10627 (15/09/2017).
	Third panel: from left to right: orbit 10641 (16/09/2017), orbit 10656 (17/09/2017).
%	Upper panel: from left to right: orbit 10584 (12/09/2017), orbit 10597 (13/09/2017), orbit 10612 (14/09/2017).
%	Lower panel: from left to right: orbit 10627 (15/09/2017), orbit 10641 (16/09/2017), orbit 10656 (17/09/2017).
\label{contours}}
\end{figure*}

We fitted the $3-40$ keV spectra of LAXPC10 instrument using the TCAF model-based {\it fits} file in {\tt XSPEC}. 
We presented the detailed results of only eleven orbits spanning our whole observation period. The accretion parameters 
did not change significantly from orbit to orbit. Since we used 2\% systematic error to fit the spectra, to consider 
the small fluctuations in the spectral fitted parameters would lead a wrong conclusion. We only presented overall variation 
of the accretion parameters during our observations. A sample LAXPC10 spectrum of orbit 10597 (observation date: 
2017 September 13) in $3-40$~keV, fitted with TCAF model is shown in Fig~\ref{tcaf_spec}. In Fig~\ref{contours}, 
we plotted confidence contours for two types of accretion rates, namely disk rate ($\dot m_d$) and halo rate ($\dot m_h$) 
for six orbits (10584, 10597, 10612, 10627, 10641, 10656) correspond to 2017 September 12, 13, 14, 15, 16, and 17. 
A clear non-correlated variation of the flow parameters is observable. In Fig.~\ref{tcaf_param}, we show the variation 
of disk rate ($\dot m_d$), halo rate ($\dot m_h$), shock location ($X_s$) and compression ratio ($R$) with day (in MJD). 
For the first observation (2017 September 12) the fitted parameters did not vary significantly. The parameters 
showed variation around MJD=58009.706 (orbit 10606). The location of shock boundary on our first observation 
(orbit 10584, MJD=58008.254) was $\sim$25.3 $r_s$ with a low compression ratio ($R=1.05$). On the fourth observation 
on MJD=58008.706 (2017 September 13), the disk rate showed a sudden increase from 1.14 to 1.69 $\dot M_{Edd}$. 
The halo rate also increased from 0.52 to 0.76 $\dot M_{Edd}$ on this observation (orbit 10606). Shock location ($X_s$) 
decreased from $\sim$23.8 to $\sim$18.7 $r_s$ on this day, but the compression ratio ($R$) remained low (1.05). 
After that, the accretion parameters remained almost constant throughout the next three observation days (till 2017 
September 17). Throughout our observation period from LAXPC data, we observed that the disk rate was high compared 
to the halo rate. The location of the shock boundary resided at a low value. These characteristics 
point to a softer state (intermediate or SS). The presence of type-C QPOs in each observation eliminates the possibility of this 
phase of the outburst in the SS. The only possible way to explain this uncommon accretion characteristic is to consider 
this period to be in the intermediate state. 

\subsection{The physical origin of the observed QPOs}
According to the shock oscillation model, the low-frequency QPOs are due to resonance oscillation of the shock (Molteni et al. 1996; 
Chakrabarti \& Manickam, 2000; Chakrabarti et al. 2015) or non-satisfaction of the Rankine-Hugoniot conditions (Ryu et al. 1997), 
which leads to unstable oscillatory shock. The resonance oscillation occurs when the infall time of the post-shock matter matches 
with the post-shock matter cooling time. To confirm whether resonance occurred or not, we computed the cooling time scales 
and infall time scales in each observation (Chakrabarti et al. 2015). 

\subsubsection{Cooling and infall time scales: Resonance Shock Oscillation} 
Following the method described in Chakrabarti et al. (2015), we calculated the cooling and infall time scales to check whether 
the observed type-C QPOs during our observation period are due to the resonance oscillation or not. According to 
Chakrabarti et al. (2015), to satisfy resonance oscillation for type-C QPOs, the ratio of the matter cooling and infall times 
should fall within roughly 50\% of unity (i.e., in between $0.5-1.5$). For our calculation, we considered the TCAF configuration 
where high angular momentum Keplerian matter ($\dot m_d$) is submerged inside the low angular momentum halo matter ($\dot m_h$). 
Depending upon the centrifugal barrier, the radially moving sub-Keplerian matter forms a shock at a location $X_s$ 
(Chakrabarti, 1989). The CENBOL is assumed to be cylindrical in shape and the total thermal energy of the CENBOL is $E_t$. 
The matter at radius $X<X_s$, moves with a velocity of $v_+$. The post-shock region cools down because of the inverse 
Comptonization of high energy electrons with the soft photons from the Keplerian disk. For supermassive black holes, cooling 
is significant due to synchrotron and bremsstrahlung processes. For the low mass X-ray binary, we only considered 
Comptonization cooling at a rate of $\Lambda_c$. After estimating the total thermal energy of the CENBOL ($E_t$) and 
cooling rate ($\Lambda_c$) due to Comptonization, we obtain the cooling time to be, 
\begin{equation}
	t_c=\frac{E_t}{\Lambda_c}.
\end{equation}
The infall time ($t_i$) of the post-shock matter moving with a velocity $v_+$, is given by,
\begin{equation}
	t_i=\frac{X_s}{v_+}.
\end{equation}

We noted the computed $t_i$, $t_c$ and their ratio ($t_c/t_i$) in Table~\ref{table2}. We see that the cooling time 
scales ($t_c$) are very low compared to the infall time scales ($t_i$). This implies that the cooling of the post-shock 
region is very efficient. Based on these results, we conclude that the origin of the QPOs is not due to resonance oscillation. 

\begin{figure*}
\vskip 0.5cm
\centering
\includegraphics[width=12cm,keepaspectratio=true]{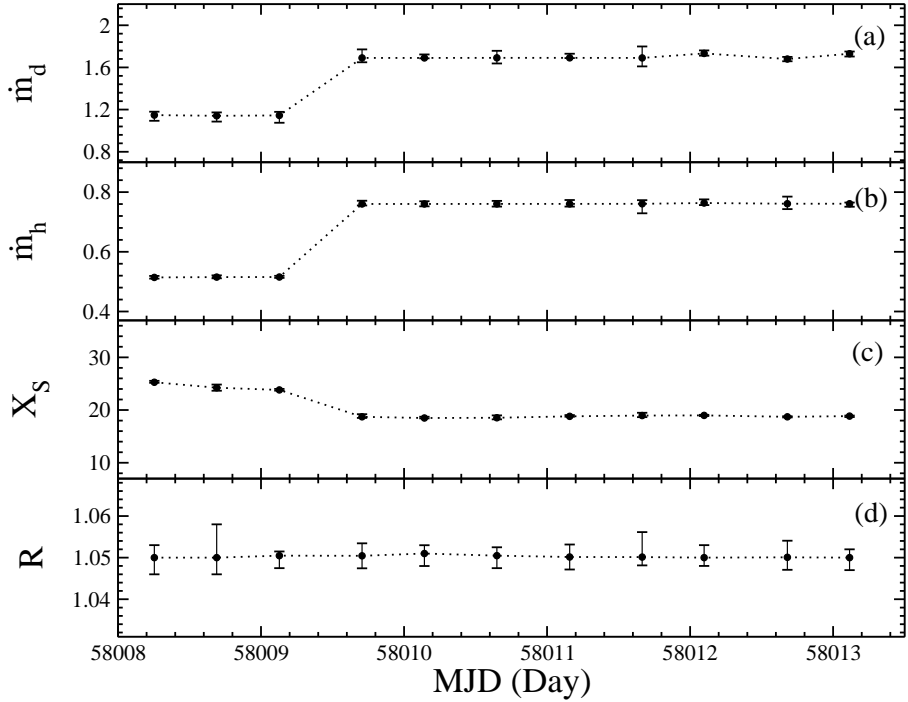}
	\caption{The variation of TCAF model fitted (a) disk rate ($\dot{m_d}$) in $\dot{M}_{Edd}$, 
	(b) halo rate ($\dot{m_h}$) in $\dot{M}_{Edd}$, (c) shock location ($X_s$) in $(r_s)$, and 
	(d) compression ratio (R) are shown with time (in MJD).
\label{tcaf_param}}
\end{figure*}

\subsubsection{Rankine-Hugoniot Conditions: Non-dissipative Shock Oscillation}
In the previous sub-Section, we have seen that the observed type-C QPOs are not due to the resonance 
oscillation since the cooling and infall time scales in the post-shock region are not found to be comparable. The other 
possible origin of these QPOs could be due to the non-satisfaction of the Rankine-Hugoniot conditions. For a steady shock, 
the Rankine-Hugoniot conditions need to be satisfied.

In Chakrabarti (1989c), the Rankine-Hugoniot shocks have been discussed in non-dissipative adiabatic flows for the hybrid 
model flow where the flow is always in vertical equilibrium. The dimensionless equations of the energy, the angular momentum, 
and the mass must be conserved to achieve a complete shock solution. The energy ($\mathscr{E}$), the pressure balance, and 
the baryon number ($M$) are given by (as in Chakrabarti 1989c),
\begin{subequations}
	\begin{align}
		\mathscr{E}_{+}&=\mathscr{E}_{-},\\
		W_{+}+\Sigma_{+}\vartheta_{+}^{2}&=W_{-}+\Sigma_{-}\vartheta_{-}^{2},\\
		\dot M_{+}&=\dot M_{-}.
	\end{align}
\end{subequations}
$W$ and $\Sigma$ denote the pressure and density integrated in vertical direction. $\vartheta$ is the non-dimensional radial 
velocity measured in the equatorial plane. The subscripts `$-$' and `$+$' implies pre- and post- shock quantities. To form 
a steady Rankine-Hugoniot shock, the satisfaction of conservation equations 3(a-c) is mandatory. During an outburst, as 
accretion rates vary, a steady shock solution is not always possible even though there are two saddle type sonic points with 
the inner sonic point having higher entropy (Chakrabarti, 1989c). In this case, one would observe a periodic movement of the 
centrifugally driven shock (Ryu et al. 1997). This oscillation would be nearly independent of the accretion rates and may also 
have some winds. This type of oscillation may be present in the intermediate state of normal outbursting sources. As shown in 
Ryu et al. (1997), the amplitude of oscillation is high, and thus QPOs could be of type-C. They reported the time period 
($\tau$ in second) of the QPOs to be in the range of
\begin{equation}
\tau \sim (4-6) \times 10^{-2} \frac{M_{BH}}{M_\odot}. 
\end{equation}
For the BHC MAXI~J1535-571 of mass $M_{BH} \sim 8.9~M_\odot$ (Shang et al. 2019), this yields a frequency in the range 
of $1.87-2.81$~Hz. This is in near-exact agreement with the observation where we observed type-C QPOs in 
the range of $1.75-2.81$~Hz.

\section{Discussions}
\label{diss}
We studied transient BHC MAXI J1535-571 during its 2017-18 outburst using {\it AstroSat} LAXPC data. We used data of
62 orbits (except operational shutdown period) from 2017 September 12 (MJD=58008.254) to 2017 September 17 (MJD=58013.155). 
We used all anode layer data of the LAXPC10 for the analysis. We studied the timing and spectral properties of the source. 
For the spectral study, we used physical accretion flow model TCAF in {\tt XSPEC}.

We observed a sharp primary QPO with a harmonic (see Table~\ref{table1}) for each of the 62 orbit data. The QPO frequencies 
were varied from $1.75$~Hz (MJD=58010.359, 2017 September 14) to $2.81$~Hz (MJD=58009.128, 2017 September 13) in a quasi-random 
manner. A sample of six continua fitted PDSs is shown in Fig.~\ref{qpo}, where the fundamental and the harmonics are also 
mentioned in the individual plots. The obtained $Q$-values and fractional $rms$ (see Table~\ref{table1}) imply these to be 
type-C QPOs. We calculated the Pearson linear correlation between the QPO frequency and the fractional rms of the 62 
observations and found a very weak negative correlation with a coefficient of r=-0.007. Signature of the low-frequency QPOs 
in MAXI~J1535-571 is also been reported by many groups (Huang et al. 2018; Stiele \& Kong 2018; Shang et al. 2019) using 
observation of other satellites as well as {\it AstroSat} (Sreehari et al. 2019; Bhargava et al. 2019). Huang et al. (2018) 
reported QPOs using the {\it Insight}-HXMT observation taken from 2017 September 6 to 23. They found different types (A, B and C) 
of QPOs in the intermediate states. They have also found a negative (soft) lag at the type-C centroid frequency and as the QPO 
frequency decreases the lag tends to zero. From these soft lags at high QPO frequencies of the type-C QPOs, they have concluded 
the system to be highly inclined. Sreehari et al. (2019) and Bhargava et al. (2019) also presented timing analysis using 
{\it AstroSat} LAXPC data and reported observation of type-C QPOs after identifying the phase of the outburst as HIMS. 
Observation of our QPO frequencies ($1.75-2.81$~Hz) roughly matches with the observed QPO 
frequencies of Sreehari et al. (2019; $1.85-2.88$~Hz) and Bhargava et al. (2019; $1.7-3$~Hz). It has been observed in many 
transient BHCs that type-C QPOs evolves monotonically in the HS and HIMS of rising and declining phases and type-B or A QPOs 
are found sporadically on and off during SIMS (see, Nandi et al. 2012; Debnath et al. 2013 and references there in). 
In Paper-I, Shang et al. (2019) also reported non-evolving type-C QPOs during our {\it AstroSat} observation period 
(12-17 Sep. 2017) in the range of $2.15-2.68$~Hz. 

To understand more detailed variation of the nature of the observed QPOs in small time scales, dynamic variation of PDSs are 
studied. Since the overall variations were not significant in a small time window of each orbit data, we made dynamic PDS by 
combining successive 13 orbits LAXPC10 data between MJD $\sim$ 58009.259 and MJD $\sim$ 58010.359, when QPO frequency varied 
in the range of $1.75$~Hz to $2.61$~Hz. A more clear variation of the primary QPO and its harmonic are observable in the 
dynamic PDS (Fig.~\ref{dynamic_pds}). This region of the outburst was an important phase. During this time, the primary dominating QPO frequency 
increased from $\sim 2.3$~Hz to $\sim 2.6$~Hz, and then decreased to $< 2.0$~Hz. Spectral analysis with 
the TCAF model, also showed a significant change in the accretion flow parameters (see, Fig. \ref{tcaf_param}).

We also studied the spectral properties of the source using the LAXPC10 data in the $3-40$~keV energy range. We used TCAF 
model-based fits file to gain insight about the accretion dynamics of the source during this period of {\it AstroSat} 
observation. From the spectral fits, we obtained accretion rates ($\dot m_d$ and $\dot m_h$) and shock parameters 
($X_s$ and $R$). The variation of TCAF model fitted parameters for eleven observations (orbits) are shown in 
Fig. \ref{tcaf_param} (also see Table~\ref{table2}). Fig~\ref{tcaf_spec} show the spectrum of orbit 10597 fitted with 
TCAF model. A high residual is present in the spectrum around 10 keV. This feature is also observed by Bhargava et al. (2019) 
and Sridhar et al. (2019). This feature was not observed in Shang et al. (2019) in the analysis of the same region of the outburst 
of the combined {\it Swift}/XRT, {\it MAXI}/GSC and {\it Swift}/BAT data using the same model (TCAF). So, we suggest that 
this is probably due to the instrumental errors rather than the intrinsic properties of the source. Although overall values 
of the obtained best-fitted flow parameters did not show significant variation, a small change in the parameters was observed 
between the third and the fourth observations. The contour plots of the two types of accretion rates ($\dot m_d$ and $\dot m_h$) 
in six orbits data shown in Fig~\ref{contours}, clearly indicates the independent nature of these two accretion components. 
The disk rate ($\dot m_d$) always remained high compared to the halo rate ($\dot m_h$). The compression ratio remained very 
low throughout our observation period and the shock was located close to the black hole ($\sim$25 $r_s$ to $\sim$19 $r_s$). 
These values indicate the signature of a softer spectral state. The presence of LFQPOs in our observational period ruled it out
as soft state. Dominating contribution of the disk rate over halo rate, the presence of weaker shock at low radius and the 
signature of the non-evolving QPOs in the PDSs confirm spectral nature of the source as intermediate state.
Due to the data quality and small variation of the accretion parameters, the precise prediction (whether HIMS or SIMS) 
of the spectral state remained inconclusive. Analyzing the broad band data of {\it Swift}/XRT, BAT and {\it MAXI}/GSC, Shang et al. (2019) concluded the 
state of the outburst as SIMS. However, the studied period of our observation was marked as HIMS by Sreehari et al. (2019) and Bhargava et al. (2019). 

Overall during our observation time (MJD = 58008.254 - 58013.155) of the {\it AstroSat}, type-C QPOs are found to vary randomly 
in a narrow range of $1.75-2.81$~Hz (see, Fig.~\ref{qpo}a and Table~\ref{table1}). This non-evolution of type-C QPOs are 
quite uncommon in a transient BHC. The origin of the type-C QPOs is still a matter of debate. Many theoretical explanations have been given and various 
models has been proposed to explain the basis of these types of QPOs. We have discussed them in Section~\ref{intro}. In this 
paper, we tried to explain the origin of the type-C QPOs observed for this source from the TCAF paradigm. In the TCAF paradigm, 
the reason behind such QPOs is believed to be the oscillation of shock boundary or CENBOL (Chakrabarti \& Manickam 2000; 
Chakrabarti et al. 2005, 2008). The shock oscillation occurs due to non-satisfaction of the Rankine-Hugoniot conditions 
(Ryu et al. 1997) or satisfaction of the resonance condition in the post-shock (CENBOL) region (Molteni, Sponholz \& 
Chakrabarti 1996, Chakrabarti et al. 2015). In the harder states, the size of the CENBOL is larger with a large shock location 
and a high compression ratio. In these states, stronger shocks create sharp shock boundaries and their oscillations produce 
type-C QPOs. As the size of the CENBOL is big, large number of soft photons from the Keplerian disk are inverse Comptonized 
to become hard photons. Due to the oscillation of the shock boundary, the CENBOL size also changes. It becomes larger when the 
shock moves away and smaller when the shock moves inward. According to the shock oscillation model, the QPO frequencies 
($\nu_{QPO}$) are inversely proportional to the infall time scales. In general, during the rising phase of an outburst of 
a transient BHC, the oscillating shock is observed to move inward with reducing $X_s$ and in the declining phase, outward 
movement of the oscillatory shock is observed. So, during the rising phase of the outburst, type-C QPO frequencies are observed 
to rise monotonically and in the declining phase monotonically decreasing nature of the QPO frequencies is observed.

To check if the origin of these type-C QPOs is due to resonance oscillation or due to non-satisfaction of the Rankine-Hugoniot 
conditions, we calculated the infall time ($t_i$) and cooling time ($t_c$) of the post-shock region. We noticed that the cooling 
time scale is of $\sim10^{3}$ order less than the infall time scale. The obtained low values of shock locations ($X_s$) from the 
TCAF model fitted spectral analysis also supports the fact that the cooling is more efficient during this period of the outburst. 
The two time scales, namely,  $t_c$ and $t_i$ were not found to be comparable. Thus we conclude that the observed type-C QPOs 
did not originate due to the resonance oscillation. The other possible origin for this type of QPOs is the shock 
oscillation due to the non-satisfaction of the Rankine-Hugoniot conditions. This happens when the flow has parameters such that 
there are two saddle type sonic points with the inner one having more entropy than the outer one, and yet, the shock condition 
is not satisfied (Chakrabarti, 1989c). This was verified by numerical simulations of Ryu et al. (1997). Here the amplitude of 
the oscillations was found to be high as to produce type-C QPOs. This oscillation would also be nearly independent of the 
accretion rates and may also have some winds produced during the rapid inward motion of the shock front. The wild infra-red 
flickering and large absorption observed for this source points to presence of outflows (Baglio et al. 2018). 
The time period of oscillation found in the simulation clearly matches with our observed frequencies. 
We also found that QPO frequencies are nearly independent of the accretion rates and vary randomly in the range of $1.75-2.81$~Hz.

Finally, we may conclude that the sharp type-C QPOs are due to the oscillations of the non-dissipating shocks when the Rankine-Hugoniot 
relation is not satisfied. Our observed frequency range ($1.75-2.81$~Hz) seems to agree with the theoretical predictions made 
($\sim 1.87-2.81$~Hz) from this consideration for MAXI~J1535-571, which has a mass of $8.9~M_\odot$.

\section{Summary}
We analyzed the {\it AstroSat} LAXPC10 data of MAXI J1535-571 from 2017 September 12 to 17. For timing analysis, we considered 
3-80 keV data of 62 orbits. We used 3-40 keV energy range data of eleven orbits for spectral study. The key findings from our 
obtained results are as follows.
\begin{enumerate}
\item The source was found to be in the intermediate spectral state during our studied period with AstroSat data. 
      High dominance of the Keplerian disk rate over the sub-Keplerian halo rate and the presence of weaker shock at low radius were observed during this phase of the outburst.
\item Dynamic PDS of continuous 13 orbit data of $\sim 1.1$~day (MJD=58009.259-58010.359) showed a small time scale variation 
      of QPOs in the range of $1.75$~Hz to $2.61$~Hz. Variation of the harmonics was also observed in the dynamic PDS.
\item For all 62 orbits data of AstroSat LAXPC, type-C LFQPOs were observed. QPOs were observed quasi-randomly in the range of 
      $1.75-2.81$~Hz. These non-evolving QPOs also signify this as intermediate spectral state. 
      We tried to find the origin of the QPOs under shock oscillation model of the TCAF paradigm. We checked whether QPOs 
      are originated due to the shock oscillation because of the satisfaction of the resonance condition or non-satisfaction of the 
      Rankine-Hugoniot conditions. Resonance oscillation was ruled out as cooling and infall time scales of the post-shock matter 
      was not matched. We found that non-satisfaction of the Rankine-Hugoniot conditions is the cause of the observed type-C QPOs. 
      A good agreement with the theory ($\sim 1.87-2.81$~Hz) in the observed frequency range ($1.75-2.81$~Hz) was also found.
\end{enumerate}

\section*{Acknowledgement}
We acknowledge the strong support from Indian Space Research Organization (ISRO)
for successful realization and operation of {\it AstroSat} mission. This work makes use of archived data
from Indian Space Science Data Centre (ISSDC). We acknowledge Prof. H. M. Antia and T. B. Katoch for their 
help and constructive suggestions regarding the data analysis and improvement of the manuscript.
D.C. and D.D. acknowledge support from DST/SERB sponsored Extra Mural Research project (EMR/2016/003918) fund.
Research of D.D. and S.K.C. is supported in part by the Higher Education Dept. of the Govt. of West Bengal, India. 
K.C. acknowledges support from DST/INSPIRE (IF170233) fellowship.
R.B. acknowledges support from CSIR-UGC NET qualified UGC fellowship (June-2018, 527223).
S.K.N., D.D. and S.K.C. acknowledge partial support from ISRO sponsored RESPOND project (ISRO/RES/2/418/17-18) fund.
H.-K. C. is supported by MOST of Taiwan under grants MOST/106-2923-M-007-002-MY3 and MOST/108-2112-M-007-003.

\clearpage

\begin{table*}
\vskip -0.2cm
%\tiny
\small
\addtolength{\tabcolsep}{-3.5pt}
\centering
\caption{Observed QPO fitted parameters}
\label{table1}
\begin{tabular}{|lcccccc||lcccccc|} 
\hline
	Orbit   & UT   & Day       &  QPO Freq.          & FWHM                    & Q   & fractional   & Orbit   & UT   & Day       &  QPO Freq.            & FWHM                    & Q   & fractional   \\
	& Date & (MJD)     &  ($\nu_{Obs}$ in Hz)  &  ($\Delta\nu$ in Hz)    &                   & rms  &        & Date & (MJD)     &  ($\nu_{Obs}$ in Hz)  &  ($\Delta\nu$ in Hz)    &     & rms  \\
(1)    &  (2) &    (3)    &       (4)             &   (5)                   &(6)  & (7)  &  (1)    &  (2) &    (3)    &       (4)             &   (5)                   &(6)  & (7)  \\
\hline
10584 &  09-12& 58008.254 &   $2.129^{+0.016}_{-0.015}$ &   $0.341^{+0.064}_{-0.085}$ &  $6.24^{+1.61}_{-1.22}$ &$9.52^{+0.048}_{-0.033}$   &  10622 &  09-14& 58010.793 &   $2.268^{+0.009}_{-0.008}$ &   $0.407^{+0.028}_{-0.024}$ &  $5.57^{+0.41}_{-0.34}$ & $10.5^{+0.025}_{-0.022}$\\  
10585 &  09-12& 58008.327 &   $2.154^{+0.006}_{-0.007}$ &   $0.318^{+0.020}_{-0.023}$ &  $6.77^{+0.53}_{-0.44}$ &$10.7^{+0.017}_{-0.015}$   &  10623 &  09-14& 58010.866 &   $2.467^{+0.009}_{-0.007}$ &   $0.437^{+0.028}_{-0.025}$ &  $5.64^{+0.38}_{-0.34}$ & $10.5^{+0.029}_{-0.026}$\\  
10586 &  09-12& 58008.401 &   $2.235^{+0.007}_{-0.007}$ &   $0.471^{+0.023}_{-0.026}$ &  $4.74^{+0.27}_{-0.24}$ &$10.8^{+0.048}_{-0.038}$   &  10624 &  09-14& 58010.937 &   $2.447^{+0.010}_{-0.010}$ &   $0.476^{+0.034}_{-0.030}$ &  $5.14^{+0.39}_{-0.35}$ & $10.6^{+0.034}_{-0.031}$\\  
10588 &  09-12& 58008.536 &   $2.482^{+0.005}_{-0.007}$ &   $0.441^{+0.018}_{-0.020}$ &  $5.62^{+0.27}_{-0.25}$ &$10.4^{+0.021}_{-0.019}$   &  10625 &  09-14& 58011.010 &   $2.181^{+0.008}_{-0.008}$ &   $0.369^{+0.030}_{-0.026}$ &  $5.91^{+0.50}_{-0.44}$ & $10.2^{+0.025}_{-0.022}$\\  
10589 &  09-12& 58008.618 &   $2.625^{+0.007}_{-0.008}$ &   $0.423^{+0.022}_{-0.024}$ &  $6.20^{+0.38}_{-0.34}$ &$10.2^{+0.031}_{-0.027}$   &  10626 &  09-14& 58011.083 &   $2.379^{+0.010}_{-0.008}$ &   $0.442^{+0.034}_{-0.031}$ &  $5.38^{+0.44}_{-0.39}$ & $10.5^{+0.028}_{-0.025}$\\  
10590 &  09-12& 58008.690 &   $2.725^{+0.007}_{-0.009}$ &   $0.449^{+0.028}_{-0.032}$ &  $6.06^{+0.46}_{-0.40}$ &$10.1^{+0.024}_{-0.021}$   &  10627 &  09-15& 58011.158 &   $2.137^{+0.008}_{-0.006}$ &   $0.377^{+0.026}_{-0.023}$ &  $5.66^{+0.41}_{-0.37}$ & $10.4^{+0.016}_{-0.014}$\\  
10592 &  09-12& 58008.762 &   $2.768^{+0.008}_{-0.009}$ &   $0.448^{+0.028}_{-0.032}$ &  $6.17^{+0.46}_{-0.40}$ &$10.2^{+0.031}_{-0.026}$   &  10628 &  09-15& 58011.230 &   $2.142^{+0.007}_{-0.005}$ &   $0.379^{+0.027}_{-0.022}$ &  $5.65^{+0.42}_{-0.35}$ & $10.3^{+0.013}_{-0.012}$\\  
10593 &  09-12& 58008.834 &   $2.634^{+0.010}_{-0.011}$ &   $0.571^{+0.030}_{-0.034}$ &  $4.61^{+0.29}_{-0.25}$ &$10.8^{+0.055}_{-0.050}$   &  10629 &  09-15& 58011.302 &   $2.122^{+0.006}_{-0.005}$ &   $0.394^{+0.023}_{-0.022}$ &  $5.38^{+0.33}_{-0.31}$ & $10.5^{+0.014}_{-0.013}$\\  
10594 &  09-12& 58008.906 &   $2.475^{+0.009}_{-0.011}$ &   $0.587^{+0.035}_{-0.038}$ &  $4.21^{+0.29}_{-0.27}$ &$10.7^{+0.038}_{-0.034}$   &  10632 &  09-15& 58011.521 &   $2.207^{+0.006}_{-0.006}$ &   $0.526^{+0.018}_{-0.016}$ &  $4.19^{+0.16}_{-0.13}$ & $11.1^{+0.028}_{-0.026}$\\  
10595 &  09-12& 58008.979 &   $2.421^{+0.009}_{-0.011}$ &   $0.483^{+0.031}_{-0.035}$ &  $5.01^{+0.38}_{-0.34}$ &$10.2^{+0.052}_{-0.044}$   &  10633 &  09-15& 58011.593 &   $2.281^{+0.006}_{-0.006}$ &   $0.362^{+0.023}_{-0.020}$ &  $6.30^{+0.42}_{-0.37}$ & $10.1^{+0.015}_{-0.014}$\\  
10596 &  09-12& 58009.051 &   $2.731^{+0.009}_{-0.011}$ &   $0.483^{+0.034}_{-0.039}$ &  $5.65^{+0.48}_{-0.42}$ &$10.1^{+0.044}_{-0.039}$   &  10635 &  09-15& 58011.665 &   $2.239^{+0.008}_{-0.008}$ &   $0.388^{+0.028}_{-0.024}$ &  $5.77^{+0.44}_{-0.39}$ & $10.6^{+0.029}_{-0.025}$\\  
10597 &  09-13& 58009.128 &   $2.811^{+0.007}_{-0.008}$ &   $0.362^{+0.020}_{-0.024}$ &  $7.76^{+0.55}_{-0.46}$ &$9.95^{+0.020}_{-0.018}$   &  10636 &  09-15& 58011.736 &   $2.254^{+0.008}_{-0.006}$ &   $0.367^{+0.028}_{-0.024}$ &  $6.14^{+0.49}_{-0.42}$ & $10.4^{+0.023}_{-0.020}$\\  
10598 &  09-13& 58009.198 &   $2.592^{+0.008}_{-0.009}$ &   $0.529^{+0.029}_{-0.032}$ &  $4.89^{+0.32}_{-0.29}$ &$10.6^{+0.031}_{-0.028}$   &  10637 &  09-15& 58011.809 &   $2.185^{+0.008}_{-0.007}$ &   $0.412^{+0.029}_{-0.025}$ &  $5.30^{+0.39}_{-0.34}$ & $10.4^{+0.103}_{-0.095}$\\  
10599 &  09-13& 58009.269 &   $2.363^{+0.006}_{-0.008}$ &   $0.470^{+0.024}_{-0.028}$ &  $5.02^{+0.32}_{-0.27}$ &$10.6^{+0.024}_{-0.022}$   &  10638 &  09-15& 58011.880 &   $2.043^{+0.009}_{-0.008}$ &   $0.355^{+0.033}_{-0.028}$ &  $5.75^{+0.56}_{-0.47}$ & $10.5^{+0.080}_{-0.078}$\\  
10600 &  09-13& 58009.343 &   $2.346^{+0.008}_{-0.009}$ &   $0.553^{+0.025}_{-0.028}$ &  $4.24^{+0.23}_{-0.20}$ &$10.9^{+0.046}_{-0.039}$   &  10639 &  09-15& 58011.953 &   $2.152^{+0.008}_{-0.006}$ &   $0.300^{+0.027}_{-0.022}$ &  $7.17^{+0.68}_{-0.56}$ & $10.0^{+0.054}_{-0.051}$\\  
10603 &  09-13& 58009.563 &   $2.622^{+0.005}_{-0.006}$ &   $0.537^{+0.016}_{-0.018}$ &  $4.88^{+0.17}_{-0.16}$ &$10.8^{+0.023}_{-0.021}$   &  10640 &  09-15& 58012.027 &   $2.218^{+0.009}_{-0.008}$ &   $0.414^{+0.031}_{-0.027}$ &  $5.35^{+0.42}_{-0.37}$ & $10.4^{+0.059}_{-0.051}$\\  
10604 &  09-13& 58009.634 &   $2.505^{+0.007}_{-0.007}$ &   $0.368^{+0.020}_{-0.024}$ &  $6.80^{+0.46}_{-0.39}$ &$9.98^{+0.019}_{-0.017}$   &  10641 &  09-16& 58012.098 &   $2.185^{+0.008}_{-0.006}$ &   $0.356^{+0.030}_{-0.026}$ &  $6.13^{+0.54}_{-0.47}$ & $10.4^{+0.026}_{-0.022}$\\  
10606 &  09-13& 58009.706 &   $2.451^{+0.008}_{-0.010}$ &   $0.441^{+0.025}_{-0.030}$ &  $5.55^{+0.40}_{-0.34}$ &$10.5^{+0.031}_{-0.028}$   &  10642 &  09-16& 58012.174 &   $2.296^{+0.007}_{-0.005}$ &   $0.376^{+0.026}_{-0.023}$ &  $6.10^{+0.44}_{-0.39}$ & $10.3^{+0.028}_{-0.026}$\\  
10607 &  09-13& 58009.778 &   $2.241^{+0.008}_{-0.009}$ &   $0.526^{+0.032}_{-0.035}$ &  $4.26^{+0.30}_{-0.27}$ &$10.6^{+0.049}_{-0.040}$   &  10643 &  09-16& 58012.245 &   $2.374^{+0.006}_{-0.005}$ &   $0.355^{+0.020}_{-0.018}$ &  $6.68^{+0.40}_{-0.36}$ & $10.2^{+0.017}_{-0.016}$\\  
10608 &  09-13& 58009.850 &   $2.030^{+0.009}_{-0.009}$ &   $0.489^{+0.029}_{-0.032}$ &  $4.15^{+0.29}_{-0.26}$ &$10.6^{+0.047}_{-0.040}$   &  10644 &  09-16& 58012.319 &   $2.609^{+0.011}_{-0.010}$ &   $0.499^{+0.030}_{-0.026}$ &  $5.22^{+0.33}_{-0.29}$ & $10.6^{+0.033}_{-0.030}$\\  
10609 &  09-13& 58009.921 &   $1.820^{+0.007}_{-0.008}$ &   $0.382^{+0.029}_{-0.031}$ &  $4.76^{+0.41}_{-0.38}$ &$10.3^{+0.017}_{-0.015}$   &  10646 &  09-16& 58012.465 &   $2.724^{+0.006}_{-0.005}$ &   $0.395^{+0.017}_{-0.014}$ &  $6.89^{+0.31}_{-0.25}$ & $10.2^{+0.052}_{-0.042}$\\  
10610 &  09-13& 58009.994 &   $1.808^{+0.007}_{-0.008}$ &   $0.399^{+0.031}_{-0.038}$ &  $4.53^{+0.45}_{-0.36}$ &$10.2^{+0.016}_{-0.014}$   &  10647 &  09-16& 58012.537 &   $2.523^{+0.012}_{-0.011}$ &   $0.605^{+0.036}_{-0.031}$ &  $4.17^{+0.27}_{-0.23}$ & $10.9^{+0.085}_{-0.077}$\\  
10611 &  09-13& 58010.067 &   $1.790^{+0.007}_{-0.008}$ &   $0.329^{+0.026}_{-0.030}$ &  $5.44^{+0.52}_{-0.45}$ &$10.1^{+0.019}_{-0.016}$   &  10648 &  09-16& 58012.609 &   $2.267^{+0.007}_{-0.006}$ &   $0.365^{+0.023}_{-0.020}$ &  $6.21^{+0.42}_{-0.37}$ & $10.2^{+0.031}_{-0.030}$\\  
10612 &  09-14& 58010.142 &   $1.813^{+0.005}_{-0.007}$ &   $0.301^{+0.020}_{-0.022}$ &  $6.02^{+0.47}_{-0.43}$ &$10.0^{+0.013}_{-0.011}$   &  10650 &  09-16& 58012.680 &   $2.352^{+0.008}_{-0.007}$ &   $0.362^{+0.022}_{-0.019}$ &  $6.49^{+0.42}_{-0.36}$ & $10.5^{+0.024}_{-0.021}$\\  
10613 &  09-14& 58010.214 &   $1.839^{+0.005}_{-0.006}$ &   $0.323^{+0.019}_{-0.022}$ &  $5.69^{+0.41}_{-0.35}$ &$10.2^{+0.013}_{-0.012}$   &  10651 &  09-16& 58012.753 &   $2.340^{+0.009}_{-0.008}$ &   $0.405^{+0.026}_{-0.023}$ &  $5.77^{+0.39}_{-0.35}$ & $10.5^{+0.043}_{-0.038}$\\  
10614 &  09-14& 58010.286 &   $1.831^{+0.005}_{-0.006}$ &   $0.331^{+0.019}_{-0.022}$ &  $5.53^{+0.40}_{-0.34}$ &$10.2^{+0.013}_{-0.012}$   &  10652 &  09-16& 58012.825 &   $2.248^{+0.008}_{-0.006}$ &   $0.402^{+0.027}_{-0.023}$ &  $5.59^{+0.40}_{-0.34}$ & $10.4^{+0.041}_{-0.039}$\\  
10615 &  09-14& 58010.359 &   $1.748^{+0.005}_{-0.006}$ &   $0.346^{+0.017}_{-0.021}$ &  $5.05^{+0.32}_{-0.27}$ &$10.4^{+0.014}_{-0.012}$   &  10653 &  09-16& 58012.896 &   $2.236^{+0.007}_{-0.006}$ &   $0.340^{+0.024}_{-0.021}$ &  $6.57^{+0.48}_{-0.43}$ & $10.2^{+0.030}_{-0.028}$\\  
10617 &  09-14& 58010.505 &   $1.937^{+0.006}_{-0.007}$ &   $0.563^{+0.020}_{-0.023}$ &  $3.44^{+0.15}_{-0.13}$ &$11.1^{+0.032}_{-0.030}$   &  10654 &  09-16& 58012.968 &   $2.141^{+0.010}_{-0.008}$ &   $0.405^{+0.030}_{-0.027}$ &  $5.28^{+0.42}_{-0.37}$ & $10.7^{+0.024}_{-0.022}$\\  
10618 &  09-14& 58010.577 &   $2.027^{+0.006}_{-0.006}$ &   $0.353^{+0.022}_{-0.024}$ &  $5.74^{+0.42}_{-0.38}$ &$10.3^{+0.012}_{-0.011}$   &  10655 &  09-16& 58013.042 &   $2.198^{+0.009}_{-0.008}$ &   $0.402^{+0.039}_{-0.031}$ &  $5.46^{+0.55}_{-0.44}$ & $10.5^{+0.034}_{-0.030}$\\  
10619 &  09-14& 58010.649 &   $1.967^{+0.005}_{-0.007}$ &   $0.354^{+0.023}_{-0.025}$ &  $5.55^{+0.41}_{-0.37}$ &$10.3^{+0.015}_{-0.013}$   &  10656 &  09-17& 58013.115 &   $2.089^{+0.007}_{-0.006}$ &   $0.366^{+0.029}_{-0.027}$ &  $5.70^{+0.47}_{-0.44}$ & $10.3^{+0.020}_{-0.019}$\\  
10621 &  09-14& 58010.721 &   $2.055^{+0.008}_{-0.009}$ &   $0.432^{+0.024}_{-0.028}$ &  $4.75^{+0.33}_{-0.28}$ &$10.5^{+0.043}_{-0.035}$   &  10657 &  09-17& 58013.155 &   $2.464^{+0.009}_{-0.008}$ &   $0.405^{+0.030}_{-0.026}$ &  $6.08^{+0.47}_{-0.42}$ & $10.7^{+0.030}_{-0.027}$\\  

\hline
\end{tabular}
\noindent{
\leftline{Col. 1 represents orbits of the observation. Columns 2 \& 3 show the UT date and MJD respectively. UT dates are in MM-DD format}
\leftline{of the year 2017. Cols. 4 \& 5 represent Lorentzian model fitted centroid frequency and full-width at half maximum (FWHM) of}
	\leftline{the observed QPOs in Hz. Cols. 6 \& 7 represent `Q (=$\nu/\Delta\nu$)' and fractional $rms$ of the QPOs.}
}
\end{table*}

\begin{table*}
%\centering
\caption{TCAF model fitted spectral parameters}
\label{table2}
\begin{tabular}{|lcc|cccc|ccc|}
\hline
Orbit& Date & MJD & $\dot{m}_d$       & $\dot{m}_h$       & $X_s$   & R  & $\chi^2$ & dof & $\chi^2_{red}$ \\
	&      &     & ($\dot{M}_{Edd}$) & ($\dot{M}_{Edd}$) & $(r_s)$ &    &    &          &   \\
 (1)   & (2) &   (3)        &  (4)    &   (5)        &  (6)              &      (7)          &  (8)    &(9)&  (10)  \\
\hline

10584&  2017-09-12&  58008.254&  $1.147^{+0.032}_{-0.054}$ &$0.514^{+0.005}_{-0.004}$ &$25.27^{+0.262}_{-0.118}$ &$1.05^{+0.003}_{-0.004}$ & 211.74& 139& 1.523\\
10590&  2017-09-12&  58008.690&  $1.141^{+0.032}_{-0.055}$ &$0.515^{+0.006}_{-0.004}$ &$24.22^{+0.600}_{-0.560}$ &$1.05^{+0.008}_{-0.004}$ & 172.91& 139& 1.240\\
10597&  2017-09-13&  58009.128&  $1.144^{+0.034}_{-0.070}$ &$0.515^{+0.004}_{-0.003}$ &$23.80^{+0.201}_{-0.134}$ &$1.05^{+0.001}_{-0.003}$ & 158.47& 139& 1.140\\
10606&  2017-09-13&  58009.706&  $1.690^{+0.081}_{-0.041}$ &$0.760^{+0.011}_{-0.005}$ &$18.72^{+0.488}_{-0.222}$ &$1.05^{+0.003}_{-0.003}$ & 164.07& 139& 1.180\\
10612&  2017-09-14&  58010.142&  $1.690^{+0.033}_{-0.004}$ &$0.760^{+0.009}_{-0.008}$ &$18.49^{+0.235}_{-0.079}$ &$1.05^{+0.002}_{-0.003}$ & 205.88& 139& 1.481\\
10619&  2017-09-14&  58010.649&  $1.690^{+0.067}_{-0.053}$ &$0.760^{+0.010}_{-0.009}$ &$18.54^{+0.437}_{-0.312}$ &$1.05^{+0.002}_{-0.003}$ & 172.17& 139& 1.238\\
10627&  2017-09-15&  58011.158&  $1.690^{+0.039}_{-0.002}$ &$0.760^{+0.013}_{-0.009}$ &$18.80^{+0.240}_{-0.074}$ &$1.05^{+0.003}_{-0.003}$ & 177.50& 139& 1.277\\
10635&  2017-09-15&  58011.665&  $1.690^{+0.109}_{-0.080}$ &$0.760^{+0.012}_{-0.032}$ &$18.95^{+0.519}_{-0.301}$ &$1.05^{+0.006}_{-0.002}$ & 152.86& 139& 1.099\\
10641&  2017-09-16&  58012.098&  $1.732^{+0.028}_{-0.021}$ &$0.763^{+0.012}_{-0.007}$ &$18.98^{+0.107}_{-0.128}$ &$1.05^{+0.003}_{-0.002}$ & 176.15& 139& 1.267\\
10650&  2017-09-16&  58012.680&  $1.679^{+0.020}_{-0.021}$ &$0.761^{+0.024}_{-0.018}$ &$18.70^{+0.116}_{-0.114}$ &$1.05^{+0.004}_{-0.003}$ & 163.48& 139& 1.176\\
10656&  2017-09-17&  58013.115&  $1.729^{+0.021}_{-0.026}$ &$0.761^{+0.004}_{-0.010}$ &$18.84^{+0.104}_{-0.172}$ &$1.05^{+0.002}_{-0.003}$ & 186.22& 139& 1.339\\

\hline
\end{tabular}
\noindent{
\leftline{$\dot{m}_d$, $\dot{m}_h$, $X_s$, $R$, $M_{BH}$ are the TCAF fitted parameters. The accretion rates ($\dot{m}_d$ and $\dot{m}_h$) are in Eddington rate.}
\leftline{$X_s$ represents shock location values in Schwarschild radius $r_s$ unit. $R$ is the compression ratio between post- and pre- shock densities. }
\leftline{TCAF model fitted $\chi^2_{red}$ values are mentioned in Cols , where `dof' represents the degrees of freedom.}
\leftline{The superscripts and subscripts are error values of 90\% confidence extracted using {\tt err} task in {\tt XSPEC}.}
\leftline{Note, the mass accretion rates are sometimes higher than one Eddington rate, even though the luminosities are sub-Eddington.}
\leftline{This is because TCAF self-consistently computes the efficiency factor $\eta$ for each data and the observed luminosity $L=\eta~L_E$.}
\leftline{The efficiency varies inversely to the shock location ($X_s$) in TCAF. So, for the above values of the accretion rates, the efficiency} 
\leftline{varies between 0.04 to 0.05.}
}
\end{table*}

\begin{table*}
%\centering
\caption{Resonance oscillation: Comparison of time scales}
\label{table3}
\begin{tabular}{|lcc|ccc|}
\hline
	Orbit& Date & MJD & $t_c$       & $t_i$       & $\frac{t_c}{t_i}$    \\
	&      &     & (sec) & (sec) &    \\
 (1)   & (2) &   (3)        &  (4)    &   (5)        &  (6)           \\
\hline

10584&  2017-09-12&  58008.254& 0.00012&   0.02098&  0.00572\\
10597&  2017-09-13&  58009.128& 0.00013&   0.02087&  0.00623\\
10612&  2017-09-14&  58010.142& 0.00010&   0.01257&  0.00796\\
10627&  2017-09-15&  58011.158& 0.00010&   0.01296&  0.00772\\
10641&  2017-09-16&  58012.098& 0.00010&   0.01308&  0.00764\\
10656&  2017-09-17&  58013.115& 0.00010&   0.01297&  0.00771\\
\hline
\end{tabular}
%\noindent{
%\leftline{Calculated values for MAXI J1535-571.}
%}
\end{table*}


\begin{thebibliography}{99}
\bibitem[Antia et al. (2017)]{Antia17} Antia, H. M., Yadav, J. S., Agrawal, P. C. et al., 2017, \apjs, 231, 10
\bibitem[Arnaud (1996)]{Arnaud96} Arnaud, K.A., 1996, ASP Conf. Ser., Astronomical Data Analysis Software and Systems V, ed. G.H. Jacoby \& J. Barnes, 101, 17
\bibitem[Bagilo et al. (2018)]{Bagilo18}Baglio, M. C., Russel, D. M., Casella, P., et al., 2018, \apj, 867, 114
\bibitem[Belloni \& Hasinger (1990)]{Belloni90} Belloni, T. \& Hasinger, G., 1990, \aap 230, 103
\bibitem[Belloni et al. (2005)]{Belloni05} Belloni, T., Homan, J., Casella, P., et al. 2005, \aap 440, 207
\bibitem[Bhargava et al. (2019)]{Bhargava19}Bhargava, Y., Belloni, T., Bhattacharya, D., \& Misra, R., 2019, \mnras, 488, 720
\bibitem[Bhattacharjee et al. (2017)]{AB17} Bhattacharjee, A., Banerjee, I., \& Banerjee, A., et al. 2017, \mnras, 466, 1372
\bibitem[Cabanac et al. (2010)]{CHP20}Cabanac, C., Henri, G., Petrucci, P.-O., et al. 2010, \mnras, 404, 738
\bibitem[Casella et al. (2005)]{Casella05} Casella, P., Belloni, T., \& Stella, L., 2005, \apj, 629, 403 
\bibitem[Chakabarti (1989a)]{C89a} Chakrabarti, S. K. 1989a, \apj, 337L, 89
\bibitem[Chakabarti (1989b)]{C89b} Chakrabarti, S. K. 1989b, \mnras, 240, 7
\bibitem[Chakabarti (1989c)]{C89c} Chakrabarti, S. K. 1989c, \apj, 347, 365
\bibitem[Chakrabarti1990]{SKC90} Chakrabarti, S. K. 1990, Theory of Transonic Astrophysical Flows (Singapore: World Scientific)
\bibitem[Chakrabarti \& Titarchuk(1995)]{CT95} Chakrabarti, S. K., \& Titarchuk, L.G., 1995, \apj, 455, 623
\bibitem[Chakrabarti(1997)]{SKC97} Chakrabarti, S.K., 1997, \apj, 484, 313
\bibitem[CM00]{CM00} Chakrabarti, S. K., \& Manickam, S. G., 2000, \apj, 531, L41
\bibitem[Chakrabarti et al. (2004)]{CAM04} Chakrabarti, S. K., Acharyya, K., \& Molteni, D., 2004, \aap, 421, 1
\bibitem[Chakrabarti et al. (2005)]{C05} Chakrabarti, S.K., Nandi, A., Debnath, D., et al. 2005, Ind. J. Phys, 79, 841 (arXiv:astro-ph/0508024)
\bibitem[Chakrabarti et al. (2008)]{SKC08} Chakrabarti, S. K., Debnath, D., Nandi, A., \& Pal, P. S., 2008, \aap, 489, L41
\bibitem[Chakrabarti et al.(2015)]{CMD15} Chakrabarti, S.K., Mondal, S., \& Debnath, D., 2015, \mnras, 452, 3451 (CMD15)
\bibitem[Chatterjee(2016)]{DC16} Chatterjee, D., Debnath, D., Chakrabarti, S. K., et al. 2016, \apj, 827, 88
\bibitem[Chatterjee  (2019)]{DC19} Chatterjee, D., Debnath, D., Jana, A., \& Chakrabarti, S. K., 2019, \apss, 364, 14
\bibitem[Chatterjee (2020)]{KC20} Chatterjee, K., Debnath, D., Chatterjee, D,. et al. 2020, \mnras, 493, 2452
\bibitem[Debnath et al. (2008)]{DD08}Debnath D., Chakrabarti S. K., Nandi A., Mandal S., 2008, Bull. Astron. Soc. India, 36, 151
\bibitem[Debnath et al.(2010)]{DD10} Debnath, D., Chakrabarti, S.K., \& Nandi, A., 2010, \aap, 520, A98
\bibitem[Debnath et al. (2013)]{DD13} Debnath, D., Chakrabarti, S. K., \& Nandi, A., 2013, AdSpR, 52, 2143
\bibitem[Debnath, Chakrabarti \& Mondal(2014)]{DD14} Debnath, D., Mondal, S., \& Chakrabarti, S. K., 2014, \mnras, 440, L121
\bibitem[Debnath, Mondal \& Chakrabarti(2015a)]{DD5a} Debnath, D., Mondal, S., \& Chakrabarti, S. K., 2015a, \mnras, 447, 1984
\bibitem[Debnath (2015b)]{DD15b} Debnath, D., Molla, A.A., Chakrabarti, S.K., \& Mondal, S., 2015b, \apj, 803, 59
\bibitem[Debnath et al (2017)]{DD17} Debnath, D., Jana, A., Chakrabarti, S. K., \& Chatterjee, D., 2017, \apj, 850, 92
\bibitem[Debnath et al (2020)]{DD20} Debnath, D., Chatterjee, D., Jana, A., Chakrabarti, S. K., \& Chatterjee, K., 2020, RAA, 20, 175
\bibitem[Dincer 2017]{Dincer17}Dincer, T., 2017, ATel, 10716
\bibitem[Huang et al. (2018)]{Huang18}Huang, Y., et al., 2018, \apj, 866, 122
\bibitem[Ingram et al. (2009)]{Ingram09}Ingram, A., Done, C., Fragile, P. C., 2009, \mnras, 397, L101
\bibitem[Jana,Debnath \& Chakrabarti(2016)]{AJ16} Jana, A., Debnath, D., Chakrabarti, S. K., et al. 2016, \apj, 819, 107
\bibitem[Jana et al. (2017)]{AJ17} Jana, A., Chakrabarti, S. K., \& Debnath, D., 2017, \apj, 850, 91
\bibitem[Jana et al. (2020)]{AJ20} Jana, A., Debnath, D., Chatterjee, D., et al. 2020, RAA, 20, 28
\bibitem[Kato \& Fukue(1980)]{Kato80}Kato, S., Fukue, J., 1980, \pasj, 32, 377
\bibitem[Kennea et al. (2017)]{Kennea17} Kennea, J. A., Evans, P. A., Beardmore, A. P., et al., 2017, ATel, 10700, 1
\bibitem[McClintock \& Remilard (2009)]{MR09} McClintock, J. E., \& Remillard, R. A. 2009, in Compact Stellar X-ray Sources (Cambridge University Press), 157–214
\bibitem[Mereminskiy \& Grebenev (2017)]{Mereminskiy17} Mereminskiy, I. A., \& Grebenev, S. A., 2017, ATel, 10734, 1
\bibitem[Miller et al. (2018)]{Miller18} Miller, J. M., Gendreau, K., Ludlam, R. M., et al., 2018, \apj, 860, L28
\bibitem[Molla et al. (2016)]{Molla16} Molla, A. A., Debnath, D., \& Chakrabarti, S. K. et al. 2016, \mnras, 460, 3163
\bibitem[Molla et al. (2017)]{Molla17} Molla, A. A., Debnath, D., \& Chakrabarti, S. K. et al. 2017, \apj, 834, 88
\bibitem[Mondal, Debnath \& Chakrabarti(2014)]{SM14} Mondal, S., Debnath, D., \& Chakrabarti, S.K., 2014, \apj, 786, 4
\bibitem[Mondal, Chakrabarti \& Debnath(2016)]{SM16} Mondal, S., Chakrabarti, S.K., \& Debnath, D., 2016, \apss, 361, 309
\bibitem[Molteni et al. (1996)]{MSC96} Molteni, D., Sponholz, H., \& Chakrabarti, S. K., 1996, \apj, 457, 805
\bibitem[Muno et al. (1999)]{MMR99}Muno, M. P., Morgan, E. H., \& Remillard, R. A., \apj, 1999, 527, 321
\bibitem[Nandi et al. (2012)]{N12} Nandi, A., Debnath, D., Mandal, S., \& Chakrabarti, S. K., 2012, \aap, 542, 56
\bibitem[Negoro et al. (2017)]{Negoro17} Negoro, H., Ishikawa, M., Ueno, et al., 2017, ATel, 10699, 1
\bibitem[Nobili et al. (2000)]{NTZB00}Nobili, L., Turolla, R., Zampieri, L., \& Belloni, T., 2000, \apj, 538, L137
\bibitem[Nowak \& Wagoner(1991)]{Nowak91}Nowak, M., Wagoner, R. V., 1991, \apj, 378, 656
\bibitem[Pahari et al. (2017)]{Pahari17} Pahari, M., Antia, H. M., Yadav, J. S., et al. 2017, \apj, 849, 16
\bibitem[Remilard \& McClintock06]{RM06} Remillard R. A., McClintock J. E., 2006, ARA\&A, 44, 49
\bibitem[Revnivstev et al. (2000)]{RTB00}Revnivtsev, M. G., Trudolyubov, S. P., Borozdin, K. N., 2000, \mnras, 312, 151
\bibitem[Russell et al. (2017)]{Russell17} Russell, T. D., Miller-Jones, J. C. A., Sivakoff, G. R., et al., 2017, ATel, 10711, 1
\bibitem[Ryu et al. (1997)]{Ryu97} Ryu, D., Chakrabarti, S. K., \& Molteni, D. 1997, \apj, 474, 378
\bibitem[Scaringi \& ASTR211 Students (2017)]{Scaringi17} Scaringi, S., \& ASTR211 Students, 2017, ATel, 10702, 1
\bibitem[Shang19]{S19} Shang, J.-R., Debnath, D., \& Chatterjee, D., et al. 2019, \apj, 875, 4 (Paper-I)
\bibitem[Shidatsu et al. (2017a)]{Shidatsu17a} Shidatsu, M., Nakahira, S., Negoro, H., et al., 2017a, ATel, 10761, 1
\bibitem[Shidatsu et al. (2017b)]{Shidatsu17b} Shidatsu, M., Nakahira, S., Negoro, H., et al., 2017b, ATel, 11020, 1
\bibitem[Shirakawa \& Lai (2002)]{SL02}Shirakawa, A., \& Lai, D., 2002, \apj, 564, 361
\bibitem[Sobczak et al. (2000)]{SMR00}Sobczak, G. J., McClintock, J. E., \& Remillard, R. A., et al. 2000, \apj, 531, 537
\bibitem[Sreehari et al. (2019)]{Sreehari19}Sreehari, H., Ravishankar, B. T., Iyer, N., et al., 2019, \mnras, 487, 928
\bibitem[Sridhar et al. (2019)]{Sridhar19}Sridhar, N., Bhattacharyya, S., Chandra, S., 2019, \mnras, 487, 4221
%\bibitem[Stella \& Vietri (1999)]{Stella99} Stella, L., Vietri, M., 1999, NuPhS, 69, 135
\bibitem[Stevens et al. (2018)]{Stevens18}Stevens, A. L., Uttley, P., Altamirano, D., et al., 2018, \apj, 865, L15
\bibitem[Stiele et al. (2013)]{Stiele13}Stiele, H., Belloni, T., Kalemci, E., et al., 2013, \mnras, 429, 2655
\bibitem[Stiele \& Kong (2018)]{SK18}Stiele, H., \& Kong, A. K. H., 2018, \apj, 868, 71
\bibitem[Tagger \& Pellat(1999)]{Tagger99} Tagger, M., \& Pellat, R. 1999, \aap, 349, 1003
\bibitem[Tagger et al.(2004)]{Tagger04} Tagger, M., Varni\'ere, P., Rodriguez, J., \& Pellat, R., 2004, \apj, 607, 410
\bibitem[Tao et al. (2018)]{Tao18} Tao, L., Chen, Y., G\"{u}ng\"{o}r, C., Huang, Y., et al., 2018, \mnras, 480, 4443
\bibitem[Tetarenko et al. (2017)]{Tetarenko17} Tetarenko, A. J., Russell, T. D., Miller-Jones, J. C. A., et al., 2017, ATel, 10745, 1
\bibitem[Titarchuk et al.(1998)]{Tit98}Titarchuk, L., Lapidus, I., Muslimov, A., 1998, \apj, 499, 315
\bibitem[Titarchuk \& Osherovich (2000)]{TO00}Titarchuk, L., \& Osherovich, V., 2000, \apj, 542, L111
\bibitem[Trudolyubov et al.(1999)]{Trudolyubov99}Trudolyubov, S., Churazov, E., \& Gilfanov, M., 1999, \aap, 351, L15
\bibitem[Verner et al. (1996)]{V96} Verner, D. A., Ferland, G. J., Korista, K. T., Yakovlev, D. G, 1996, \apj, 465, 487
\bibitem[Vignarca et al. (2003)]{VMB03}Vignarca, F., Migliari, S., Belloni, T., et al. 2003, \aap, 397, 729
\bibitem[Wilms et al. (2000)]{WAM00}Wilms, J., Allen, A., \& McCray, R., 2000, \apj, 542, 914
\bibitem[Xu et al. (2017)]{Xu17} Xu, Y., Harrison, F. A., Garcia, J. A., et al., 2017, \apj, 852, L34
\end{thebibliography}
\end{document}